\documentclass[aps,pra,reprint,amsmath,amssymb,superscriptaddress,nofootinbib,floatfix]{revtex4-2}
\usepackage{graphicx}
\graphicspath{{figures/}{./}}
\usepackage{xcolor}
\usepackage{booktabs}
\usepackage{array}
\usepackage[hidelinks]{hyperref}
\usepackage{dsfont}

\newcommand{\Qtot}{Q_{\mathrm{tot}}}
\newcommand{\Hphys}{\mathcal{H}_{\mathrm{phys}}}
\newcommand{\nq}{n_q}
\newcommand{\wpass}{w_{\mathrm{pass}}}
\newcommand{\weff}{w_{\mathrm{eff}}}
\newcommand{\Uxy}{U^{xy}}
\newcommand{\id}{\mathds{1}}
\providecommand{\ket}[1]{\lvert #1\rangle}

\providecommand{\expval}[1]{\langle #1\rangle}

\begin{document}

\title{Disentangling Expressibility, Symmetry Protection, and Hardware Noise in Variational Quantum Simulation of the Two-Flavor Schwinger Model}

\author{Karthikeya Machiraju}
\email{karthikeyamachiraju005@gmail.com}
\affiliation{Department of Computer Science and Engineering (AI \& ML), PES University, Bengaluru 560085, India}

\author{Krishna Sujith}
\affiliation{Department of Electronics and Communication Engineering, PES University, Bengaluru 560085, India}

\author{Kaustav Bhowmick}
\email{kaustavbhowmick@pes.edu}
\affiliation{Department of Electronics and Communication Engineering, PES University, Bengaluru 560085, India}

\date{\today}

\begin{abstract}
Existing quantum simulations of the two-flavor Schwinger model have run at a
single lattice size, and it is not known how far the variational approach can
be pushed or which of its known weaknesses stops it first. We map that
boundary. Following the two-flavor Schwinger model from $N=2$ to $6$ staggered
lattice sites, we find that the binding constraint at reachable sizes is
hardware noise rather than circuit expressibility or trainability, and we
identify $N=3$ as the immediately viable extension of existing trapped-ion
experiments. The Variational Quantum Eigensolver (VQE) is a leading near-term
route to lattice gauge theories in regimes where Monte Carlo sampling fails,
but its behavior as the lattice grows is set by three mechanisms that are
usually studied in isolation. We treat expressibility, symmetry protection, and
hardware noise in a single convention, and we use the half-system
entanglement entropy to locate where expressibility and trainability bind.
The energy error of a charge-conserving ansatz collapses onto one function of
the ratio $p/d$ of the number of variational parameters to the dimension of the
physical sector, and falls by more than two orders of magnitude as $p/d$
rises through order unity. The collapse gives the expressibility condition
$L(4N-1)\gtrsim\binom{2N}{N}$, where $L$ is the number of circuit layers. The
condition is local in the chemical potential: at $N=3$ the layer count that
suffices at zero chemical potential leaves a $74.38\%$ error near the
first-order boundary, and no number of restarts recovers it, while one further
layer reaches $0.08\%$ at a cost of $7.80$ percentage points of cumulative
infidelity.
Charge conservation gives a double protection. When the qubit count doubles
from $4$ to $8$, the gradient variance of the constrained ansatz, normalized by the size of
the Hamiltonian, falls to $1/3.56$ of its starting value, against $1/13.57$ for
an unconstrained circuit
under the same protocol, and the constrained ansatz is immune to a
charge-sector leakage that is otherwise detectable only through
$\expval{\Qtot^2}$.
Hardware noise is modeled in two ways, as a global contraction applied to a
converged state and as per-gate local channels inside the optimization loop.
The two disagree, and a fixed-parameter control shows why: in the same local
channel, a noiselessly optimized state and an in-loop optimized one suppress
the first-order slope discontinuity by statistically indistinguishable
amounts, with a median of $9.79\%$ against $16.16\%$, while the global
contraction predicts $26.77\%$ at $26.00\%$ cumulative infidelity. The noise
model, not the optimization protocol, sets the size of the degradation. The next lattice size,
$N=4$, is noise-limited at $p=1.00\%$: a noiseless control reaches $0.12\%$ mean error where
the same circuit at $1.00\%$ depolarizing reaches $25.69$ to $52.81\%$.
\end{abstract}


\maketitle

\section{Introduction}
Lattice gauge theories (LGTs) are the primary nonperturbative framework for
quantum field theory. Their standard computational tool is Euclidean Monte
Carlo, yet several physically important regimes remain difficult for
importance sampling: finite chemical potential, real-time evolution, and a
topological $\theta$-term. In these regimes the path-integral measure becomes
oscillatory and importance sampling fails~\cite{bauer2023,banuls2020}. The
failure of importance sampling is not a limitation of classical computation as
a whole; classical tensor-network methods can reach parts of these regimes at
small volume~\cite{schwaegerl2025,barata2025}. The obstruction is nonetheless
severe, because the sign problem is NP-hard in
general~\cite{troyerwiese2005}. These are not edge cases: they include the
dense-matter phase diagram, real-time string breaking, and out-of-equilibrium
dynamics. The present work addresses how far one quantum-computational route
around this obstruction, the variational preparation of LGT ground states, can
be pushed before it stops working, and which of its known weaknesses stops it
first.

The Variational Quantum Eigensolver (VQE)~\cite{peruzzo2014,tilly2022,cerezo2021}
is the leading near-term route around the sign problem. It works in the
Hamiltonian formulation, where the operator stays Hermitian with a real
spectrum at every chemical potential~\cite{banuls2020}. It also suits noisy
intermediate-scale quantum (NISQ) hardware, through shallow circuits and a
hybrid quantum-classical loop. The case for it rests on one premise that has
not been tested: that VQE keeps working as the lattice grows toward sizes a
classical computer cannot follow. Testing that premise, and establishing how
the method degrades once it fails, is the contribution of the present work.

The two-flavor Schwinger model, quantum electrodynamics in $1{+}1$
dimensions~\cite{coleman1976}, is the natural place to test the premise: it
confines, has a mass gap, breaks chiral symmetry~\cite{dempsey2022}, and
undergoes chemical-potential-driven first-order transitions~\cite{banuls2017},
while staying small enough to check against exact diagonalization. Throughout
this paper $N$ denotes the number of staggered lattice sites and $F$ the number
of fermion flavors. We work at $F=2$, so a lattice of $N$ sites is encoded in
$\nq=2N$ qubits. What the literature does not yet provide is the scaling
answer. Melzer \textit{et al.}~\cite{melzer2025} have just demonstrated
two-flavor VQE on a four-qubit trapped-ion processor, following earlier
Schwinger-model dynamics experiments~\cite{martinez2016,klco2018}. Rosanowski
\textit{et al.}~\cite{rosanowski2026} have since taken the same
gauge-invariant, Gauss-law-enforcing construction to two flavors of
$(2{+}1)$-dimensional QED at finite density on a $2\times2$ lattice. Both are
important proofs that the construction runs on hardware, and both work at a
single fixed system size; what neither can reveal is whether VQE scales, or
which obstacle binds first as it stops scaling. That is the question we ask
here, for the same ansatz family, in one spatial dimension where exact
diagonalization remains available as a control.

Classical matrix-product-state
calculations~\cite{schwaegerl2025,barata2025} give precise energy and spectrum
benchmarks, yet they face their own entanglement-growth barrier in the
real-time and finite-density regimes that motivate quantum simulation. They fix
the frontier a quantum method must cross; they are not a way to avoid crossing
it. Scalable circuits fixed classically have prepared the one-flavor vacuum on
$100$ qubits~\cite{farrell2024}, and a three-flavor model at finite chemical
potential has been studied variationally with proof-of-principle hardware
runs~\cite{schuster2024}, but for two flavors anyone planning the step beyond
$N=2$ has no map of where VQE fails. Both ways of guessing wrong are expensive: aiming
too large returns noise-dominated nonsense, and aiming too small wastes the
device. A scaling study must ask not only whether VQE fails, but which
mechanism binds first. Three mechanisms can do so independently, and we treat
them in that order throughout this work:
(i)~\emph{Expressibility}. A circuit with too few parameters spans only a
submanifold of the physical Hilbert space. VQE then returns the best state
that submanifold contains, not the true ground state.
(ii)~\emph{Trainability}. Gradient variance can vanish exponentially with
system size, a barren plateau~\cite{mcclean2018}. Training a variational
circuit is also NP-hard in general~\cite{bittel2021}.
(iii)~\emph{Hardware noise}. Decoherence biases the measured energy and can
flatten the landscape further~\cite{wang2021}. Error mitigation removes part
of that bias, but introduces limits of its own. Each of the three mechanisms
has been studied, but in isolation and on different models and
ans\"atze: symmetry-preserving circuits~\cite{anselmetti2021,gard2020} and
barren-plateau diagnostics~\cite{mcclean2018,larocca2022} were developed
largely on their own terms.

Treating the three mechanisms one at a time is standard, but it is not
sufficient here. Expressibility and hardware noise are both controlled by the
number of layers $L$ in the ansatz circuit, and they pull in opposite
directions. A study of expressibility alone prescribes a larger $L$, a study of
noise alone prescribes a smaller $L$, and neither can determine whether any $L$
satisfies both at a given lattice size $N$. We therefore follow one theory
across system sizes, in one convention and one codebase, and ask which obstacle
binds first and at what $N$ the method stops being usable. Treating the three
in a single convention is what makes the comparison possible at all: it allows
the obstacle that binds to be named, and the lattice size at which it binds to
be quoted, neither of which follows from three separate single-obstacle
studies.

We present what is, to our knowledge, the first study of a multi-flavor
lattice gauge theory to compare all three obstacles in a single convention: we follow
two-flavor Schwinger VQE from $N=2$ to $N=6$ and separate the three obstacles
well enough to determine which of them limits the method first. The central
result is a single design rule. Here $L$ denotes the number of layers in the
ansatz circuit, $p=L(4N-1)$ the number of variational parameters it carries,
and $d=\binom{2N}{N}$ the dimension of the charge-neutral sector that contains
the ground state. Across every lattice size $N$ and every layer count $L$
tested, the relative energy error of the charge-conserving ansatz is set by the
single ratio $p/d$, and falls sharply once $p/d$ reaches unity.

The upper end of the range, $N=6$, is set by the same comparison and not by
hardware. Over $N=2$ to $6$ the sector dimension $d$ grows from $6$ to $924$,
while the smallest layer count that satisfies $p\ge d$ grows from $1$ to $41$.
The range therefore brackets the $p/d\approx1$ crossover at every depth we
test. Against this backdrop, the paper makes four contributions, each of which
goes beyond what counting arguments, or single-obstacle studies, already give.

\begin{itemize}
\item \emph{Expressibility collapse.} We show that the accuracy of the
simulation is controlled by a single ratio: the number of tunable parameters in
the circuit, divided by the size of the physical space that the circuit has to
cover. Once that ratio reaches one, the energy error drops by more than two
orders of magnitude, and the drop happens at $p/d$ of order one for every lattice
size we tested. The practical value is that choosing a circuit depth stops
being a trial-and-error exercise and becomes a one-line check
(Section~\ref{sec:express}).

\item \emph{Double symmetry protection.} We measure what enforcing charge
conservation inside the circuit actually buys, by running the same problem with
and without it. It keeps the training signal from flattening out as the system
grows, and it stops the optimizer from drifting into states that are not
physical and reporting an energy lower than the true one. We also give a cheap
measurement that detects the second failure whenever it occurs
(Section~\ref{sec:symmetry}).

\item \emph{An entanglement diagnostic.} We show that one quantity, the
entanglement across the middle of the lattice, explains why a given circuit
depth is forced across the central phase and why the optimization is slow
there, while the flavor-sector structure, not entanglement, explains where the
ansatz fails near the transition (Section~\ref{sec:entangle}).

\item \emph{A trapped-ion noise budget.} We give the measurement cost, the
extent to which gate noise smears out the phase transition, and the point at
which the standard error-mitigation method stops working, together with
simulated rather than extrapolated estimates for $N=4$
(Section~\ref{sec:noise}).
\end{itemize}

Two conventions are used throughout. The first is the chemical-potential
assignment $\nu_0=2\sqrt{x}\,K$ on flavor $0$ with $\nu_1=0$, which matches the
experimental benchmark~\cite{melzer2025}; the symbols $\nu_f$, $x$ and $K$ are
defined in Section~\ref{sec:model}. The second is the kinetic
Hamiltonian of Section~\ref{sec:model}. All numbers reported in this paper come
from a single statevector and density-matrix codebase, and the
exact-diagonalization benchmarks reproduce the values of~\cite{melzer2025}
exactly at two of the three published chemical potentials and to $0.136$ energy
units at the third, which is $0.06\%$ of the energy range spanned by the three
phases. Section~\ref{sec:verify} gives the point-by-point comparison.

\section{Model and Physical Sector}\label{sec:model}
This section defines the model and the notation used in the rest of the paper.
Section~\ref{sec:qubitham} maps the continuum theory to a qubit Hamiltonian and
defines every parameter that appears later. Section~\ref{sec:sector} identifies
the physical subspace in which all subsequent calculations are performed, and
gives its dimension. Section~\ref{sec:verify} checks the mapping against the
published values of~\cite{melzer2025}.

\subsection{From continuum QED to a qubit Hamiltonian}\label{sec:qubitham}
A variational quantum algorithm acts on a finite register of qubits, so the
continuum field theory has to be reduced to a finite spin Hamiltonian before
any of it can be run. Three reductions are needed, and each removes one obstacle
to that goal: the continuum has to be discretized without introducing spurious
fermion species, the gauge field has to be removed so that only matter degrees
of freedom remain, and the remaining fermionic operators have to be rewritten
as spin operators that a qubit register can hold. The continuum theory is
therefore mapped to qubits in three steps:
(a)~the Kogut--Susskind staggered formulation~\cite{kogut1975} places particles
on even and antiparticles on odd lattice sites, which avoids fermion doubling
and gives one fermionic mode $\phi_{n,f}$ for each site $n=1,\dots,N$ and each
flavor $f$;
(b)~in $1{+}1$ dimensions Gauss's law $L_n-L_{n-1}=Q_n$, relating the electric
field $L_n$ on link $n$ to the charge $Q_n$ on site $n$, is solved exactly,
which eliminates the gauge field in favor of the long-range charge-charge
interaction $\sum_n L_n^2 \to \sum_n(\sum_{k\le n}Q_k)^2$ that encodes
confinement;
(c)~the Jordan--Wigner transformation~\cite{jordan1928}
$\phi_n^\dagger\mapsto(\prod_{k<n}(-iZ_k))\sigma_n^+$ assigns the qubit index
$p=nF+f$ to each mode, where $F=2$ is the number of flavors. The resulting
dimensionless spin Hamiltonian is
\begin{align}
W = &-x\sum_{n,f}\tfrac{1}{2}\!\left(X_{n,f}Z_{\mathrm{str}}X_{n+1,f}
+ Y_{n,f}Z_{\mathrm{str}}Y_{n+1,f}\right)\nonumber\\
&+\sum_{n,f}\tfrac{\nu_f}{2}\left(Z_{n,f}+1\right)
+\sum_{n}\Big(\textstyle\sum_{k\le n}Q_k\Big)^{2}.
\label{eq:W}
\end{align}

The three terms of Eq.~\eqref{eq:W} are the kinetic, chemical-potential, and
electric terms, respectively. The coupling $x=1/(ag)^2$ gets its value from the
lattice spacing $a$ and the gauge coupling $g$, and is fixed at $x=16$
throughout this paper, which is the value used in the trapped-ion benchmark of
Ref.~\cite{melzer2025} and therefore the value at which our results can be
checked against published numbers. Nothing in the analysis that follows is
specific to $x=16$; the dependence of the results on $x$ is listed among the
open cases in the Limitations of Section~\ref{sec:discussion}. The quantity
$\nu_f=2\sqrt{x}\,\kappa_f/g$ is the dimensionless chemical potential of flavor
$f$, with $\kappa_f$ the corresponding dimensionful chemical potential; it is
the parameter that controls the filling of each flavor, and it is the only
parameter varied in the sweeps of Sections~\ref{sec:symmetry}
and~\ref{sec:noise}. We use a single dimensionless control parameter $K$,
defined by $\nu_0=2\sqrt{x}\,K$ with $\nu_1=0$, so that $K$ is the chemical
potential applied to flavor $0$. We work at zero bare fermion mass,
$m_0=m_1=0$, as in~\cite{melzer2025}, so Eq.~\eqref{eq:W} carries no mass term.
The Jordan--Wigner string of step~(c)~\cite{jordan1928} is
$Z_{\mathrm{str}}=Z_{p_1+1}\cdots Z_{p_2-1}$, where $p_1$ and $p_2$ are the
qubit indices of the two modes being coupled.

The smallest case in which all three terms of Eq.~\eqref{eq:W} are nontrivial
is $N=2$ sites at $F=2$ flavors, which is also the size realized in the
trapped-ion experiment of Ref.~\cite{melzer2025} and the size at which
Section~\ref{sec:verify} checks the convention. It is therefore written out
explicitly. For $N=2$ and $F=2$, which is $\nq=4$ qubits, the three terms
reduce to the Pauli strings
\begin{align}
W_{\mathrm{kin}}&=-\tfrac{x}{2}(X_0Z_1X_2+Y_0Z_1Y_2+X_1Z_2X_3+Y_1Z_2Y_3),\\
W_{\mathrm{chem}}&=\tfrac{\nu_0}{2}\big[(Z_0{+}1)+(Z_2{+}1)\big],
\label{eq:Wchem}\\
W_{\mathrm{elec}}&=\tfrac{3}{2}\id+Z_0+Z_1+\tfrac{1}{2}Z_0Z_1,
\label{eq:Welec}
\end{align}
with $\nu_0=2\sqrt{x}\,K$ as above. Both diagonal terms of Eq.~\eqref{eq:W}
count occupation through $(\id+Z_q)/2$, so that a set bit denotes an empty
mode; Eq.~\eqref{eq:Welec} follows from squaring the resulting cumulative
charge and is the form against which the implementation is checked.

One detail of the mapping is easily got wrong, and it is a disadvantage of the
staggered formulation rather than a feature of it. The chemical-potential term,
written explicitly in Eq.~\eqref{eq:Wchem}, carries no staggered $(-1)^n$
factor: the two sites enter with the same sign. Only a nonzero fermion mass
would produce the alternating sign, and the mass is zero here. Had the
staggered sign been applied to $\nu_f$, Eq.~\eqref{eq:Wchem} would instead read
$\tfrac{\nu_0}{2}[(Z_0{+}1)-(Z_2{+}1)]$, which reproduces the correct energy at
$K=0$, where the whole term vanishes, but is wrong at every $K\ne0$. The effect
of the wrong choice is not small. It displaces the first-order transitions in
$K$, and therefore misplaces the phase boundary at which
Section~\ref{sec:entangle} locates the collapse of the entanglement entropy,
and at which Section~\ref{sec:noise} measures the noise-induced suppression of
the energy slope discontinuity. Both of those results are quoted at a specific
value of $K$, so both would be reported at the wrong chemical potential under
the staggered convention. We therefore flag the convention pitfall explicitly,
and confirm the unstaggered form against all three benchmark points
of~\cite{melzer2025} in Section~\ref{sec:verify}.

\subsection{Physical sector and phase structure}\label{sec:sector}
The total charge $\Qtot=\sum_q(\id+Z_q)/2-\nq/2$ commutes with $W$ (see
Eq.~\eqref{eq:W}). Physical states therefore obey \emph{global} charge
neutrality $\Qtot\ket{\psi}=0$, which is the single residual constraint left
once the \emph{local} Gauss law has been solved to eliminate the gauge links.
We write $\Hphys$ for the subspace of charge-neutral states, and call it the
physical sector. Every energy, gradient and fidelity reported in this paper
refers to $\Hphys$ unless stated otherwise. Its dimension is
\begin{equation}
d(N)=\binom{2N}{N},\qquad d=6,\,20,\,70,\,252,\,924
\label{eq:d}
\end{equation}
for $N=2,\dots,6$.

The number of fermions of each flavor, $N_f$, also commutes with $W$, so
$\Hphys$ splits into superselection sectors labeled by the set
$\mathbf{N}=(N_0,N_1)$, with piecewise-linear energies
$E_{\mathbf N}(\nu_0)=\nu_0 N_0+E^{\min}_{\mathbf N}$. Only the flavor-$0$
number $N_0$ appears explicitly, because $\nu_1=0$ in the convention of
Section~\ref{sec:qubitham}; $N_0$ is the quantity plotted in
Fig.~\ref{fig:symmetry}(d) and used to label the branches of
Table~\ref{tab:protocol}. Since the sector energy rises with $N_0$ at fixed
$\nu_0>0$, the favored sector is the one of lowest flavor-$0$ occupation, so
$N_0$ steps \emph{downward} as $K$ increases. First-order transitions occur at
level crossings between these sectors. The sector energies are exactly linear in $K$, with slope $8N_0$ at $x=16$, so
the crossings follow in closed form. The terminal crossings lie at
$\lvert K_{\mathrm{crit}}\rvert=3.95$, $5.57$ and $6.36$ at $N=2$, $3$ and $4$,
respectively; the $N=2$ value agrees with the $\pm3.96$ quoted
in~\cite{melzer2025} to the precision given there. At odd $N$ a further
crossing lies at $K=0$, where $\nu_0=\nu_1=0$ makes the two flavor-exchanged
sectors degenerate. At every $N$ there is more than
one crossing, and we quote throughout the terminal crossing, beyond which the
flavor-$0$ occupation has reached zero and no longer changes.

\begin{figure*}[t]
\centering
\includegraphics[width=\textwidth]{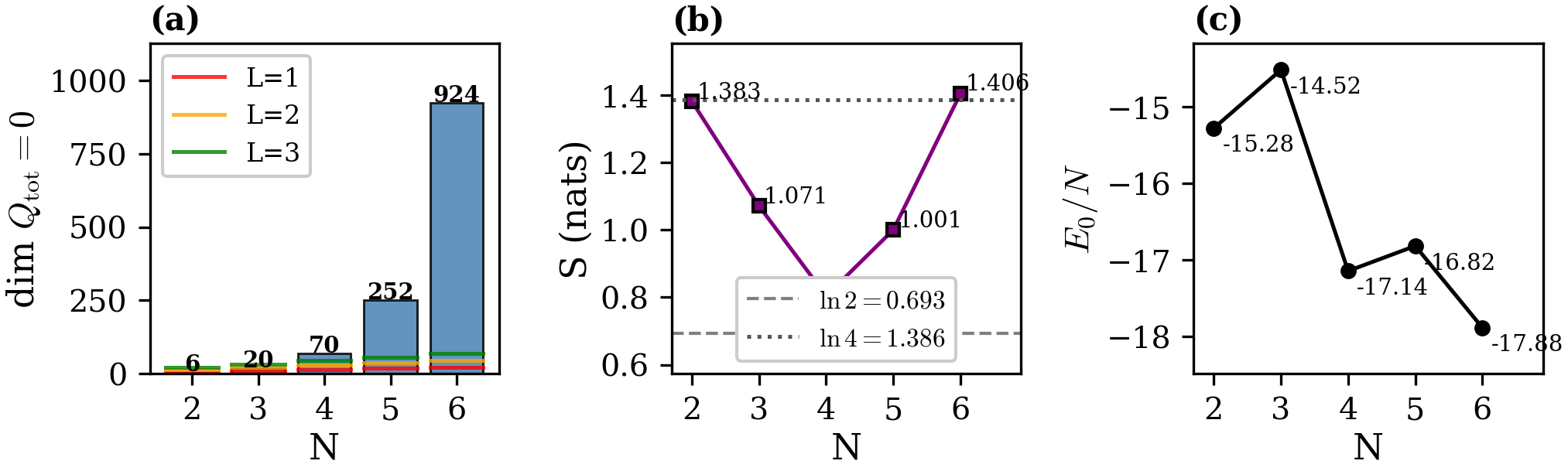}
\caption{Exact-diagonalization baseline ($K=0$, $x=16$, $F=2$), computed in
this work. (a)~Physical-sector dimension $d=\binom{2N}{N}$ (bars) against the
$L$-layer parameter budgets $p=L(4N-1)$ (dashed); the $L=1$ budget is overtaken
at $N=3$, the $L=2$ budget at $N=4$. (b)~Half-system entanglement entropy,
exceeding $\ln2$, that is one ebit, for every $N$. (c)~Energy per site,
showing even-$N$/odd-$N$ staggering of the finite lattice.}
\label{fig:foundation}
\end{figure*}

The dimension $d$ in Eq.~\eqref{eq:d} grows combinatorially with $N$, while the
parameter budget of any fixed-depth circuit grows only linearly in $N$. The two
must therefore cross, and where they cross is what sets the expressibility
problem. Figure~\ref{fig:foundation}(a) shows the two quantities together: the
sector dimension as bars, and the parameter budgets $p=L(4N-1)$ of the first
three layer counts as dashed lines. The $L=1$ budget is overtaken at $N=3$ and
the $L=2$ budget at $N=4$, so a circuit of fixed depth ceases to have enough
parameters to cover the physical sector at a lattice size that can be read off
directly. Locating that crossing is the subject of Section~\ref{sec:express}.

\subsection{Hamiltonian verification}\label{sec:verify}
The staggering convention of Section~\ref{sec:qubitham} changes the energy at
every $K\ne0$, so the mapping must be checked against published values before
it is used at larger $N$. We therefore evaluate the spin Hamiltonian of
Eq.~\eqref{eq:W} at the three chemical-potential points reported
in~\cite{melzer2025}, with the results shown in Table~\ref{tab:verify}. At
$N=2$ we obtain $-223.000$, $-30.5644$ and $+1.000$ at $K=-14$, $0$ and $+10$,
respectively, in the same order of $K$ as the rows of Table~\ref{tab:verify}.
Two of the three values agree exactly with the published ones. At $K=0$ a
previously reported ground-state energy of $-30.700$ units~\cite{melzer2025} is
close to the presently calculated $-30.5644$ units. The difference of $0.136$
units cannot be a diagonalization tolerance, since both values are exact
diagonalizations of the same six-dimensional sector. The same work quotes
$-30.500$ units for this quantity from its own statevector simulation, a value
that lies on the opposite side of ours from its tabulated one, and we cannot
identify the source of the difference from the published information. The difference is not physically
significant: it is $0.06\%$ of the $224$-unit range that $W$ spans across the
three phases, and no conclusion drawn in this paper depends on the $K=0$ energy
to better than one unit. We report the difference rather than absorbing it into
a tolerance, because the chemical-potential term of Eq.~\eqref{eq:W} vanishes
identically at $K=0$, so at that point no convention choice, staggered or
otherwise, distinguishes the two calculations, while the ground states at the
two nonzero points are polarized product states that test only the diagonal
terms. The $K=0$ point is the only one of the three that tests the hopping term. All energies
are in the dimensionless units of Eq.~\eqref{eq:W} and are not normalized by
the number of sites.

\begin{table}[b]
\caption{Hamiltonian verification at $N=2$ and $x=16$, where $x$ is the
coupling defined in Eq.~\eqref{eq:W}. Ground-state energies $E_0$ obtained in
this work are compared against the two values published
in~\cite{melzer2025}, its tabulated exact value and the result of its own
statevector simulation. $\lvert\Delta\rvert$ is taken against the tabulated
exact value. The $K=0$ difference of $0.136$ units is $0.06\%$ of the
$224$-unit energy range spanned by the three phases. Energies are quoted to
three decimal places throughout the paper; the four-decimal value at $K=0$ is
retained here because the comparison is made at that precision.}
\label{tab:verify}
\centering
\begin{tabular}{crrrr}
\toprule
$K$ & $E_0$ (this work) & $E_0$~\cite{melzer2025} & $E_0$~\cite{melzer2025} & $\lvert\Delta\rvert$\\
    &                   & (exact)                 & (statevector)           & \\
\midrule
$-14$ & $-223.000$  & $-223.000$ & ---       & $0.000$\\
$0$   & $-30.5644$  & $-30.700$  & $-30.500$ & $0.136$\\
$+10$ & $+1.000$    & $+1.000$   & ---       & $0.000$\\
\bottomrule
\end{tabular}
\end{table}

Table~\ref{tab:exact} lists the ground-state energies $E_0$, the per-site
energies $E_0/N$, and the half-system entanglement entropies $S_{A|B}$ for
every $N$ studied, all obtained in this work by exact diagonalization in the
physical sector, using the convention verified in Table~\ref{tab:verify}. No
published two-flavor values exist for $N\ge3$, so the comparison
against~\cite{melzer2025} is confined to Table~\ref{tab:verify}. The entropy
shown in Fig.~\ref{fig:foundation}(b) is non-monotone in $N$, dipping to
$0.804$~nats at $N=4$, but it exceeds $\ln2=0.693$~nats for every $N$. Every
ground state therefore carries more than one ebit across the half-system cut,
where one ebit is the entanglement of a maximally entangled pair of qubits and
equals $\ln2$ nats. No product-state ansatz can represent such a state exactly. At odd $N$ and
$K=0$ the ground state is doubly degenerate between the sectors
$(N_0,N_1)=((N+1)/2,(N-1)/2)$ and its flavor mirror, whose half-system
entropies differ, $1.071$ and $1.163$ at $N=3$ and $1.001$ and $1.127$ at $N=5$;
the table quotes one member of each pair, so part of the alternation between
even and odd $N$ reflects this degeneracy. Fidelities at these points are
computed against the projector onto the ground space, which changes every
reported value by less than $0.01$.
Figure~\ref{fig:foundation}(c) shows the corresponding energies per site.
$E_0/N$ alternates between even and odd $N$ rather than approaching its limit
monotonically, which is the even-odd staggering of the finite Kogut--Susskind
lattice and not a physical effect. It is the reason the finite-size fit of
Section~\ref{sec:entangle} is performed over all five lattice sizes rather than
over the two largest.

\begin{table}[t]
\caption{Exact diagonalization in the physical sector ($K=0$, $x=16$), computed
in this work. Energies are in the dimensionless units of Eq.~\eqref{eq:W}.}
\label{tab:exact}
\centering
\begin{tabular}{cccrr}
\toprule
$N$ & $d=\binom{2N}{N}$ & $E_0$ & $E_0/N$ & $S_{A|B}$ (nats)\\
\midrule
2 & 6   & $-30.564$  & $-15.282$ & $1.383$\\
3 & 20  & $-43.566$  & $-14.522$ & $1.071$\\
4 & 70  & $-68.576$  & $-17.144$ & $0.804$\\
5 & 252 & $-84.090$  & $-16.818$ & $1.001$\\
6 & 924 & $-107.303$ & $-17.884$ & $1.406$\\
\bottomrule
\end{tabular}
\end{table}

Two results of this section carry forward. The physical sector has dimension
$d=\binom{2N}{N}$, which is the target a circuit must cover, and every ground
state in it is entangled across the half-system cut. Section~\ref{sec:express}
asks how deep a circuit has to be before it can reach such a state.

\section{Obstacle I: Expressibility}\label{sec:express}
Expressibility is the first of the three obstacles to scaling VQE identified in
Section~\ref{sec:model}. The question it poses is whether a circuit of a given
depth can represent the ground state at all, independently of how well the
optimizer is able to find it. Expressibility is quantified here by the
parameter count $p=L(4N-1)$ of the circuit relative to the dimension
$d=\binom{2N}{N}$ of the physical sector, and by nothing else; $L$ is the
number of circuit layers and $N$, as in Section~\ref{sec:model}, the number of
staggered lattice sites. Section~\ref{sec:ansatz} defines the ansatz and counts
its parameters. Section~\ref{sec:condition} converts that count into the
expressibility condition. Section~\ref{sec:collapse} tests the condition
against VQE results from $N=2$ to $N=6$.

\subsection{Multi-layer charge-conserving ansatz}\label{sec:ansatz}
A single circuit layer is sufficient at $N=2$ but not beyond, so we generalize
the single-layer fermionic-exchange circuit introduced in~\cite{schuster2024}
and used in~\cite{melzer2025,anselmetti2021}
to $L$ stacked circuit layers. Stacking layers is the one knob that raises the
parameter count while leaving the charge symmetry of the circuit intact, which
is what allows expressibility to be studied separately from the symmetry
protection of Section~\ref{sec:symmetry}. Each layer applies the
gauge-invariant exchange gate
$\Uxy_{ij}(\theta)=e^{-i\theta(X_iX_j+Y_iY_j)/2}$ on each of the $(\nq-1)$
neighbouring qubit pairs $(i,j)$, followed by $R_z$ rotations on all $\nq$
qubits. Every gate conserves the excitation count, since $X_iX_j+Y_iY_j$
commutes with $Z_i+Z_j$. The state therefore remains inside the physical sector
$\Hphys$ of Section~\ref{sec:sector} throughout the optimization.

Two orderings of the $(\nq-1)$ exchange gates within a layer are in use, and
they carry the same parameter count and the same charge symmetry. In the
construction of Ref.~\cite{melzer2025}, the even-indexed pairs are applied
first and the odd-indexed pairs second, as a brick wall; we apply the pairs
sequentially along the chain, $(0,1),(1,2),\dots$. The two are not equivalent
at $L=1$, and Section~\ref{sec:collapse} treats that difference as a result
rather than a detail. All the results henceforth use the sequential ordering
unless stated otherwise. The parameter count per layer is $2\nq-1=4N-1$, so an
$L$-layer circuit has $p=L(4N-1)$ variational parameters in total. Each $\Uxy$
gate decomposes into two CNOTs and single-qubit rotations, which fixes the
hardware cost quoted in Table~\ref{tab:express}. The ansatz is therefore
specified by the single integer $L$, and what remains is to decide how large
$L$ must be, which is the subject of Section~\ref{sec:condition}.

\subsection{Expressibility condition}\label{sec:condition}
A pure state in $\Hphys$ traces out a manifold of real dimension $2(d-1)$,
while an $L$-layer ansatz with $p=L(4N-1)$ parameters covers a submanifold of
real dimension at most $p$. Strict coverage of the full manifold would require
$p\gtrsim 2(d-1)$. The variational target, however, is a single ground state
rather than the entire manifold, and we find numerically in
Section~\ref{sec:collapse} that the energy-error collapse sets in near the
weaker ratio $p\approx d$. As a numerically motivated heuristic we therefore
propose the expressibility condition
\begin{equation}
L(4N-1)\ge\binom{2N}{N},\qquad
L_{\min}(N)=\left\lceil\frac{\binom{2N}{N}}{4N-1}\right\rceil ,
\label{eq:express}
\end{equation}
and refer to Eq.~\eqref{eq:express} by that name throughout the rest of the
paper. The expressibility condition is a necessary scaling guide, not a
rigorous sufficiency bound. Parameter count alone does not fix which states a
circuit can reach; the gate ordering and the qubit connectivity matter as well.
The $N=2$, $L=1$ case discussed in Section~\ref{sec:collapse} is an explicit
counterexample.

Two hardware costs follow directly from $L_{\min}$. The decomposed two-qubit
circuit depth is $D=4L_{\min}+4N-6$ for the sequential ordering used here,
against about $4L_{\min}$ for a brick wall, and the CNOT count is
$2L_{\min}(2N-1)$.
Table~\ref{tab:express} evaluates both for every $N$. The expressibility
condition is so far only a counting argument. Whether it actually predicts the
point at which VQE starts to work is a question that the numerical results can
answer, and it is tested in Section~\ref{sec:collapse}.

\begin{table}[b]
\caption{Expressibility and hardware requirements, evaluated from
Eq.~\eqref{eq:express}.}
\label{tab:express}
\centering
\begin{tabular}{cccccc}
\toprule
$N$ & $\nq$ & $d$ & params/$L$ & $L_{\min}$ & depth $D$\\
\midrule
2 & 4  & 6   & 7  & 1  & 6\\
3 & 6  & 20  & 11 & 2  & 14\\
4 & 8  & 70  & 15 & 5  & 30\\
5 & 10 & 252 & 19 & 14 & 70\\
6 & 12 & 924 & 23 & 41 & 182\\
\bottomrule
\end{tabular}
\end{table}

\subsection{Data collapse at $p/d\approx1$}\label{sec:collapse}
\begin{figure*}[t]
\centering
\includegraphics[width=\textwidth]{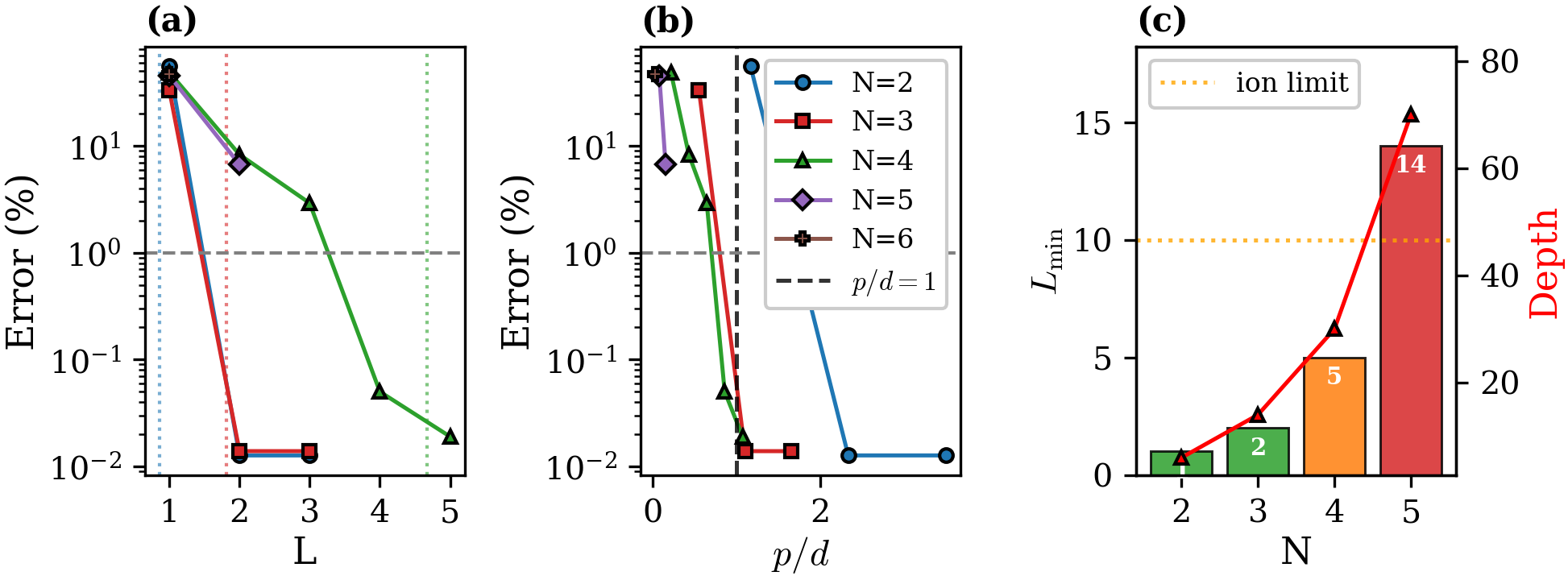}
\caption{Expressibility. (a)~Relative energy error versus ansatz layers $L$;
dotted verticals mark $L_{\min}$. The crossings for $N=5$ and $N=6$ lie at
$L=14$ and $L=41$ and are outside the plotted range. (b)~The same data collapse
onto a single function of $p/d$, with the error transition at $p/d\approx1$.
(c)~Minimum-layer two-qubit depth $4L_{\min}+4N-6$ for $N=2$ to $5$, color-coded by
trapped-ion feasibility (green: feasible; orange: marginal; red: infeasible).
The dashed horizontal line in panels (a) and (b) marks $1\%$ relative error.
Symbol and color coding in panel (a) is shared with panel (b).}
\label{fig:express}
\end{figure*}

The main result of this subsection is that the control parameter for VQE
accuracy is the ratio $p/d$ of parameter count to sector dimension, and not $N$
or $L$ separately. All runs use the derivative-free COBYLA
optimizer~\cite{powell1994} with up to eight random restarts.
Table~\ref{tab:vqe} lists the resulting energies and relative errors for every
$(N,L)$ pair tested. Figure~\ref{fig:express}(a) shows the same relative errors
against the layer count $L$, with the dotted verticals marking $L_{\min}$ from
Table~\ref{tab:express}. The error falls steeply once $L$ reaches $L_{\min}$ at
each $N$, but the four curves do not lie on top of one another, so $L$ by
itself is not the control parameter. Figure~\ref{fig:express}(b) plots the same
errors against $p/d$ instead. Runs at different $N$ and different $L$ now fall
on a single curve, and the error drops by more than two orders of magnitude as
$p/d$ rises through order unity, between about $0.7$ and $2$.

The collapse has a simple geometric reading. The parameter count $p$ is the
number of independent directions along which the ansatz can move, and the
sector dimension $d$ is the size of the space it must cover. Their ratio
therefore measures whether the variational manifold is large enough to contain
the ground state, and unity is the point at which it first can be. The collapse
is not implied by the counting bound of Eq.~\eqref{eq:express}, which fixes a
threshold for each $(N,L)$ pair separately and says nothing about runs at
different $N$ falling on a common curve. It is a regularity of the numerical
data, and it is what makes $p/d$ usable as a design parameter.

The change in accuracy across $p/d=1$ is abrupt rather than gradual, and the
size of the jump is set out here. At $N=3$, the $L=1$ circuit ($p/d=0.55$)
gives $28.81\%$ error, while $L=2$ ($p/d=1.10$) gives $0.01\%$. The same
crossing occurs at $N=4$, where the energies were obtained by statevector VQE
in the full Hilbert space and compared against the exact charge-neutral ground
state: $L=3$ ($p/d=0.64$) leaves $2.91\%$, $L=4$ ($p/d=0.86$) leaves $0.05\%$,
and $L=5$ ($p/d=1.07$) leaves $0.02\%$, so at $N=4$ most of the drop is already
complete just below unity. The value $L_{\min}=5$ is therefore
sufficient for near-exact energies at $N=4$. At $N\ge5$ with $L\le2$ the error
is at least $6.69\%$. Points on both sides of $p/d=1$ are available at $N=2$, $3$
and $4$, as Fig.~\ref{fig:express}(b) shows. The $N=5$ and $N=6$ entries of
Table~\ref{tab:vqe} sit at $p/d=0.15$ and $0.03$ and therefore probe only the
undercomplete side, since Eq.~\eqref{eq:express} places their crossings at
$L=14$ and $L=41$. Those two crossings lie beyond the layer range plotted in
Fig.~\ref{fig:express}(a) and beyond the depths we ran, which is why they do
not appear in the figure. The collapse is thus demonstrated across three system
sizes and is consistent with, but not established by, the two largest.

Figure~\ref{fig:express}(c) converts $L_{\min}$ into the decomposed two-qubit
circuit depth $D=4L_{\min}+4N-6$ of Table~\ref{tab:express}, color-coded by
trapped-ion feasibility. A depth is taken as feasible when it lies within the
coherence budget of current trapped-ion devices, marginal when it is comparable
to that budget, and infeasible when it exceeds it. On that basis $N=2$ and
$N=3$, at depths $6$ and $14$, are feasible; $N=4$, at depth $30$, is marginal;
and $N\ge5$, at depth $70$ and above, is beyond NISQ reach.

The expressibility condition of Eq.~\eqref{eq:express} is necessary but not
sufficient, and the gate ordering of Section~\ref{sec:ansatz} is what makes it
so. At $N=2$ and $L=1$ the seven-parameter circuit is overcomplete by count
($p/d=1.17$), yet under the sequential ordering it yields $49.26\%$ error,
because that ordering generates a Lie-group orbit inside $\Hphys$, under the
gate set of Section~\ref{sec:ansatz}, that does not contain the ground state.
Under the brick-wall ordering the same seven parameters behave differently: it
has been reported~\cite{melzer2025} that a single-layer circuit, evaluated on a
statevector simulator at the VQE-optimal parameters, reproduces the exact
energies to within roughly $1.00\%$ at all three chemical potentials. The two
results are not in conflict. The likely explanation is that the two orderings
generate different reachable sets at equal $p/d$; we have not proved this, and
the controlled brick-wall against sequential comparison at fixed $p/d$ that
would establish it is listed among the open cases in
Section~\ref{sec:discussion}. A second layer closes the gap under either
ordering. Connectivity and ordering therefore matter as much as parameter count
whenever $p/d$ is near unity, which is the practical content of
Eq.~\eqref{eq:express} being necessary rather than sufficient.

Restart statistics separate expressibility failure from optimizer failure. At
fixed $N=3$ over $20$ random restarts, the undercomplete $L=1$ manifold
converges deterministically to a single wrong minimum, with zero variance
across restarts. The marginally overcomplete $L=2$ ansatz ($p/d=1.10$) instead
recovers the true ground state in roughly three-quarters of restarts, so $5$ to
$8$ restarts suffice in practice. Rapid convergence on an undercomplete
manifold is therefore a signal of exhausted variational freedom, not of
success, and diagnostic checks against exact diagonalization are indispensable
whenever $p/d<1$. Expressibility at zero chemical potential is therefore a
solved design question at the sizes studied: once $p/d\ge1$ and the
connectivity has been checked, the ansatz can represent the ground state.
Whether the same layer count suffices elsewhere in the phase diagram is a
separate question, taken up in Section~\ref{sec:boundary_express}.

\begin{table}[b]
\caption{Multi-layer VQE scaling ($K=0$, $x=16$), sequential intra-layer gate
ordering. Energies are in the dimensionless units of Eq.~\eqref{eq:W}. OC:
overcomplete ($p\ge d$); UC: undercomplete.}
\label{tab:vqe}
\centering
\begin{tabular}{ccccrl}
\toprule
$N$ & $L$ & params & $p/d$ & $E_{\mathrm{VQE}}$ & err.\ (\%)\\
\midrule
2 & 1 & 7  & 1.17 & $-15.508$ & 49.26 (OC$^\dagger$)\\
2 & 2 & 14 & 2.33 & $-30.561$ & 0.01 (OC)\\
3 & 1 & 11 & 0.55 & $-31.016$ & 28.81 (UC)\\
3 & 2 & 22 & 1.10 & $-43.560$ & 0.01 (OC)\\
3 & 3 & 33 & 1.65 & $-43.560$ & 0.01 (OC)\\
4 & 1 & 15 & 0.21 & $-35.060$ & 48.88 (UC)\\
4 & 2 & 30 & 0.43 & $-62.889$ & 8.29 (UC)\\
4 & 3 & 45 & 0.64 & $-66.580$ & 2.91 (UC)\\
4 & 4 & 60 & 0.86 & $-68.542$ & 0.05 (UC)\\
4 & 5 & 75 & 1.07 & $-68.564$ & 0.02 (OC)\\
5 & 2 & 38 & 0.15 & $-78.462$ & 6.69 (UC)\\
6 & 1 & 23 & 0.03 & $-57.156$ & 46.73 (UC)\\
\bottomrule
\end{tabular}
\\[2pt]
{\footnotesize $^\dagger$Overcomplete by count but wrong ordering at $L=1$; see
the discussion of gate ordering in Section~\ref{sec:collapse}.}
\end{table}

\subsection{The threshold depends on the chemical potential}\label{sec:boundary_express}
Sections~\ref{sec:condition} and~\ref{sec:collapse} established the
expressibility condition and the collapse at $K=0$. The condition is a
statement about the ground state at that point, and the ground state changes
character across the sweep, so the sufficient layer count need not be the same
everywhere. We test this at $N=3$, where Eq.~\eqref{eq:express} gives
$L_{\min}=2$ and $p/d=1.10$. Near the first-order boundary the $L=2$ ansatz
fails outright. Table~\ref{tab:boundary_L} lists the noiseless result at three
chemical potentials inside the window $K\in[4.0,5.0]$, obtained with $32$
random restarts at a maximum of $4000$ iterations per restart. At $K=4.5$ the
ansatz returns $-1.938$ against the exact value $-7.566$, a relative error of
$74.38\%$, with a fidelity of $0.61$ against the exact ground state.

The failure is one of representation and not of optimization. Increasing the
restart budget by a factor of four and the iteration budget by a factor of two
changes the recovered energy by $0.0001$ energy units. Twenty-three of the $32$
restarts land within $0.007$ of the same wrong value, which is the
zero-variance signature already identified for the undercomplete $L=1$ manifold
in Section~\ref{sec:collapse}. A third layer removes the failure completely: at
$K=4.5$ the $L=3$ ansatz reaches $-7.560$, a relative error of $0.08\%$ with a
fidelity of $0.9999$, which rounds to the $1.00$ entered in
Table~\ref{tab:boundary_L}, and a fourth layer returns the identical value, so
$L=3$ saturates.

\begin{table}[b]
\caption{Noiseless VQE near the $N=3$ first-order boundary, $x=16$, $32$
restarts at $L=2$ and $8$ restarts at $L=3$ and $L=4$. The $L=2$ ansatz
satisfies $p/d=1.10$ yet fails; the failure is removed by one additional
layer. Energies are in the dimensionless units of Eq.~\eqref{eq:W}.}
\label{tab:boundary_L}
\centering
\begin{tabular}{ccrrrc}
\toprule
$K$ & $L$ & $p/d$ & $E_{\mathrm{VQE}}$ & err.\ (\%) & fidelity\\
\midrule
$4.0$ & 2 & 1.10 & $-5.169$ & $55.31$  & $0.68$\\
$4.5$ & 2 & 1.10 & $-1.938$ & $74.38$  & $0.61$\\
$5.0$ & 2 & 1.10 & $+0.535$ & $114.99$ & $0.38$\\
$4.5$ & 3 & 1.65 & $-7.560$ & $0.08$   & $1.00$\\
$4.5$ & 4 & 2.20 & $-7.560$ & $0.08$   & $1.00$\\
\bottomrule
\end{tabular}
\end{table}

Two consequences follow: (a)~Eq.~\eqref{eq:express} should be read as a
condition evaluated at a specific point in the phase diagram, so that one layer
is added to $L_{\min}$ in the boundary region whenever the chemical potential
is scanned; and (b)~the counterexample in Section~\ref{sec:collapse}, where
$N=2$ at $L=1$ satisfies $p/d=1.17$ and still fails, is not isolated. Both
cases show that $p/d\ge1$ is a necessary condition, and that the sufficiency of
a given layer count must be checked against the state actually being prepared.

The extra layer is not free. Each layer adds $2(2N-1)$ CNOTs, so the move from
$L=2$ to $L=3$ at $N=3$ raises the CNOT count from $20$ to $30$ and the
cumulative infidelity at a per-CNOT depolarizing probability of $1.00\%$ from
$18.20\%$ to $26.00\%$. Expressibility and hardware noise therefore make
opposite demands on the same parameter in the same region of the phase diagram,
which is the situation Section~\ref{sec:noise} quantifies. Expressibility is
therefore a design question with two answers, one at zero chemical potential
and a stricter one near the boundary. Both leave open whether the optimizer can
find the state the manifold now contains, which is the trainability question
taken up in Section~\ref{sec:symmetry}.

\section{Obstacle II: Symmetry Protection}\label{sec:symmetry}
The second of the three obstacles to scaling VQE, listed in
Section~\ref{sec:model}, is trainability. Even an ansatz that can represent the
ground state is useless if the optimization landscape is flat or if the
optimizer is free to wander into states that are not physical. This section
shows that imposing charge conservation addresses both problems at once.
Section~\ref{sec:plateau} measures the gradient variance.
Section~\ref{sec:leakage} shows what goes wrong when the symmetry is not
imposed. Section~\ref{sec:penalty} prices the standard alternative, a soft
penalty term. Section~\ref{sec:stability} compares the stability of the two
optimizations.

\subsection{Suppression of barren plateaus under charge conservation}\label{sec:plateau}
For a generic random $L$-layer circuit, the variance of the energy gradient
scales as $2^{-\nq}$~\cite{mcclean2018}. To test whether the charge-conserving
ansatz shows the same suppression, we sample $20$ random parameter vectors
$\theta\sim\mathcal{U}[0,2\pi)^p$ at $L=2$ and estimate
$\mathrm{Var}[\partial_\mu E]$ across the sample, using central finite
differences with step $\delta=10^{-3}$ and averaging over the parameter index
$\mu$. Energies are in the dimensionless units of Eq.~\eqref{eq:W}, as in
Table~\ref{tab:vqe}, and are not normalized.

The observed variances are $24.15$, $23.02$ and $21.19$ at $N=2$, $3$ and $4$,
respectively, a total decay by a factor $1.14$ as $\nq$ doubles from $4$ to
$8$. Applying the identical protocol to the unconstrained hardware-efficient
circuit of Section~\ref{sec:leakage} gives $10.47$, $5.35$ and $2.41$, a decay
by a factor $4.34$ over the same range. The operator $W$ itself grows with $N$, so we normalize each variance by
$\sum_ic_i^2=\mathrm{Tr}(W^2)/2^{\nq}$, which is $260.5$, $522.5$ and $814.5$ at
$N=2$, $3$ and $4$. The normalized decay factors are $3.56$ for the constrained
ansatz and $13.57$ for the unconstrained one, against reference factors of
$4.00$ and $16.00$ for variances decaying as $2^{-\nq/2}$ and $2^{-\nq}$ over the
same range. The constrained ansatz therefore decays at close to the half rate
and the unconstrained one at close to the full rate, and their ratio of $3.81$
is the same with or without normalization. Shallow circuits on four to eight
qubits are not expected to show a barren plateau, and three sizes cannot
establish one; what the comparison establishes is that the constrained ansatz
decays more slowly than the unconstrained one by nearly a factor of four.
Whether the difference persists asymptotically is a question for the
dynamical-Lie-algebra framework of Ref.~\cite{ragone2024}, since the physical
sector itself grows exponentially. The mechanism is the symmetry
constraint. The charge-conserving ansatz operates entirely within $\Hphys$, of
dimension $d=\binom{2N}{N}\ll2^{\nq}$, so the barren-plateau exponent scales as
$\log_2 d$ rather than $\nq$~\cite{larocca2022}. We define the protection ratio
as
\begin{equation}
\frac{\log_2 d}{\nq}=\frac{\log_2\binom{2N}{N}}{2N}
\xrightarrow{N\gg1}1-\frac{\log_2(\pi N)}{4N},
\label{eq:protection}
\end{equation}
which is the effective fraction of the qubit register that the optimization
actually explores.

Equation~\eqref{eq:protection} gives $0.65$, $0.72$ and $0.77$ at $N=2$, $3$
and $4$, respectively, so the effective system size at $N=4$ is about $6.13$
qubits rather than $8$. Charge conservation therefore slows gradient decay, and vanishing gradients are not the binding constraint
at the sizes studied. The same principle, that explicitly conserving gauge
charge shrinks the effective problem dimension, also underlies recent classical
tensor-network methods using virtual rishons~\cite{rogerson2026}. There the
reduction lowers the classical contraction cost, whereas here it acts on the
trainability of a variational quantum circuit.

Flat gradients are not the only way an optimization can fail, however.
Section~\ref{sec:leakage} shows a second and more damaging failure mode that
charge conservation also removes.

\subsection{Charge-sector leakage}\label{sec:leakage}
\begin{figure*}[t]
\centering
\includegraphics[width=\textwidth]{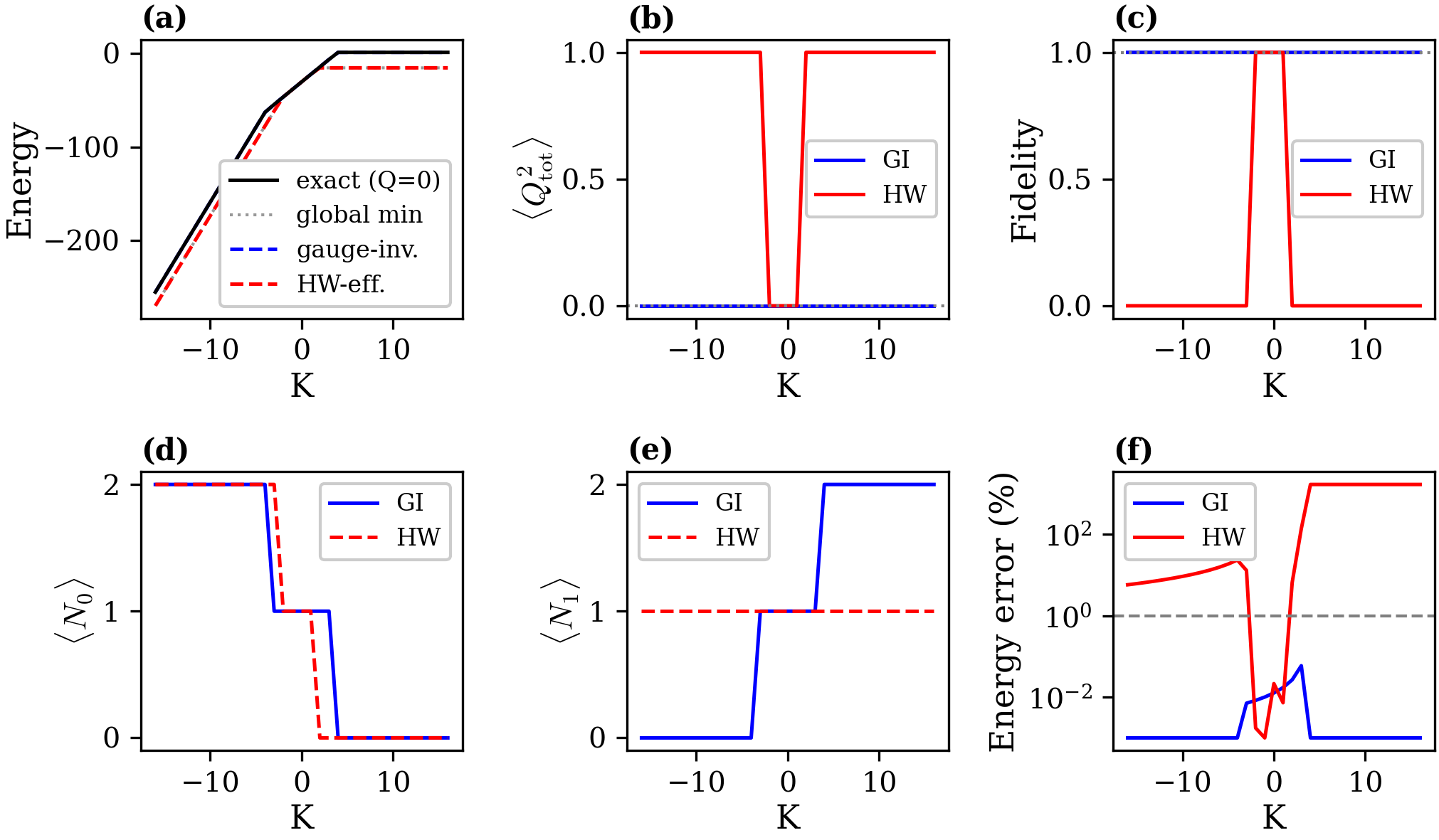}
\caption{Gauge-invariant (GI) vs.\ hardware-efficient (HW) ansatz across the
chemical potential at $N=2$, $L=2$ (33-point sweep). (a)~Ground-state energy:
the HW result drops below the physical ground state outside the window
$K\in[-2,1]$ (charge-sector leakage), while GI tracks it, and the
global minimum over all charge sectors is shown for reference.
(b)~Global-charge violation $\expval{\Qtot^2}$: identically zero for GI, and
unity for HW outside the window $K\in[-2,1]$. (c)~Fidelity with the physical ground state:
unity for GI, collapsing to zero for HW. (d)~Flavor-0 particle number,
integer-valued and stepping down from $2$ to $0$ as $K$ increases for GI,
displaced for HW. (e)~Flavor-1 particle number, showing the complementary
upward step for GI while HW remains pinned at unity.
(f)~Relative energy error against the $\Qtot=0$ ground state: GI stays below
$10^{-2}\%$ across the sweep, HW exceeds $10\%$ beyond the boundary.}
\label{fig:symmetry}
\end{figure*}

Symmetry protection has a second and independent benefit. A hardware-efficient
(HW) ansatz~\cite{kandala2017}, built from $R_yR_z$ rotations and a CNOT ladder
and carrying no charge constraint, searches the full $2^{\nq}$-dimensional
space. Only a fraction $\binom{2N}{N}/2^{2N}\approx1/\sqrt{\pi N}$ of that
space is physical, which is $37.50\%$ at $N=2$ and $31.25\%$ at $N=3$. At large
$\lvert K\rvert$ the chemical potential makes the nonphysical charge sectors
energetically favorable, and the unconstrained optimizer finds them. On the
$33$-point sweep of Fig.~\ref{fig:symmetry} the HW ansatz remains in the
physical sector only for $K\in[-2,1]$, with the transition falling between grid
points at $K=-3$ and $K=-2$ and between $K=1$ and $K=2$; the $1$-unit grid does
not resolve it more finely. That window is narrower than the first-order
boundary at $K_{\mathrm{crit}}\approx\pm3.95$, so leakage sets in well before
the physical transition, and it is not symmetric in $K$ because the convention
$\nu_1=0$ of Section~\ref{sec:qubitham} breaks the symmetry between the two
flavors.
Table~\ref{tab:gihw} compares the gauge-invariant (GI) and HW ans\"atze at
$N=2$. At $K=-14$ the physical ground state is $-223.000$, but the global
minimum over all charge sectors lies in the $\Qtot=+1$ sector at $-239.500$,
and the HW ansatz at $L=2$ converges there with fidelity $0.00$. At $K=+10$ it
returns $-13.570$ against the physical value $+1.000$, an absolute error of
$14.570$ energy units. The corresponding relative error of $1457.00\%$ is large
only because the denominator is: the physical energy at $K=+10$ is $+1.000$
units, which is $0.45\%$ of the $224$-unit range that $W$ spans across the
three phases, so the physical energy happens to lie close to zero on the scale
of the Hamiltonian even though the state itself is unremarkable. The absolute
error is the meaningful figure at that point, and the same caution applies to
the $K=+10$ row of Table~\ref{tab:shot}.

Without measuring $\expval{\Qtot^2}$~\cite{kokail2019} these results would be
indistinguishable from success. The GI ansatz, confined to $\Hphys$ by
construction, gives $\expval{\Qtot^2}=0.00$ and integer particle numbers at
every $K$; see Fig.~\ref{fig:symmetry}(a)--(e). It also recovers the sharp
first-order transitions at $K_{\mathrm{crit}}\approx\pm3.95$. Leakage can also
be suppressed without changing the ansatz, by adding a penalty term to the cost
function. Section~\ref{sec:penalty} shows what that costs.

\begin{table}[b]
\caption{Gauge-invariant (GI) vs.\ hardware-efficient (HW) ansatz, $L=2$,
$N=2$, $x=16$. Errors against the physical ($\Qtot=0$) ground state. The large
relative error at $K=+10$ reflects the small physical $\lvert E\rvert$; the
convention-independent signature of leakage is $\expval{\Qtot^2}\ne0$ with zero
fidelity, not the percentage magnitude.}
\label{tab:gihw}
\centering
\begin{tabular}{cccrrc}
\toprule
$K$ & ansatz & $E_{\mathrm{VQE}}$ & err.\ (\%) & $\expval{\Qtot^2}$ & fidelity\\
\midrule
$-14$ & GI & $-223.000$ & $0.00$    & $0.00$ & $1.00$\\
$-14$ & HW & $-239.500$ & $7.40$    & $1.00$ & $0.00$\\
$0$   & GI & $-30.560$  & $0.01$    & $0.00$ & $0.99$\\
$0$   & HW & $-30.560$  & $0.01$    & $0.00$ & $0.99$\\
$+10$ & GI & $+1.000$   & $0.00$    & $0.00$ & $1.00$\\
$+10$ & HW & $-13.570$  & $1457.00$ & $1.00$ & $0.00$\\
\bottomrule
\end{tabular}
\end{table}

\subsection{Penalty terms and their cost}\label{sec:penalty}
A soft penalty $C(\theta)=\expval{W}+\lambda\expval{\Qtot^2}$, with penalty
strength $\lambda$, can suppress the leakage of amplitude into the unphysical
charge sectors. The penalty method itself is standard~\cite{stannigel2014}; the
threshold at which it starts to work is derived here. Requiring the penalized
cost of the nonphysical global minimum to exceed that of the physical state, we
obtain the crossover condition
\begin{equation}
\lambda\ge\lvert E_{\Qtot=0}-E_{\mathrm{global}}\rvert ,
\label{eq:lambda}
\end{equation}
where $E_{\Qtot=0}$ is the physical ground-state energy and
$E_{\mathrm{global}}$ is the lowest energy over all charge sectors.

\begin{figure*}[t]
\centering
\includegraphics[width=\textwidth]{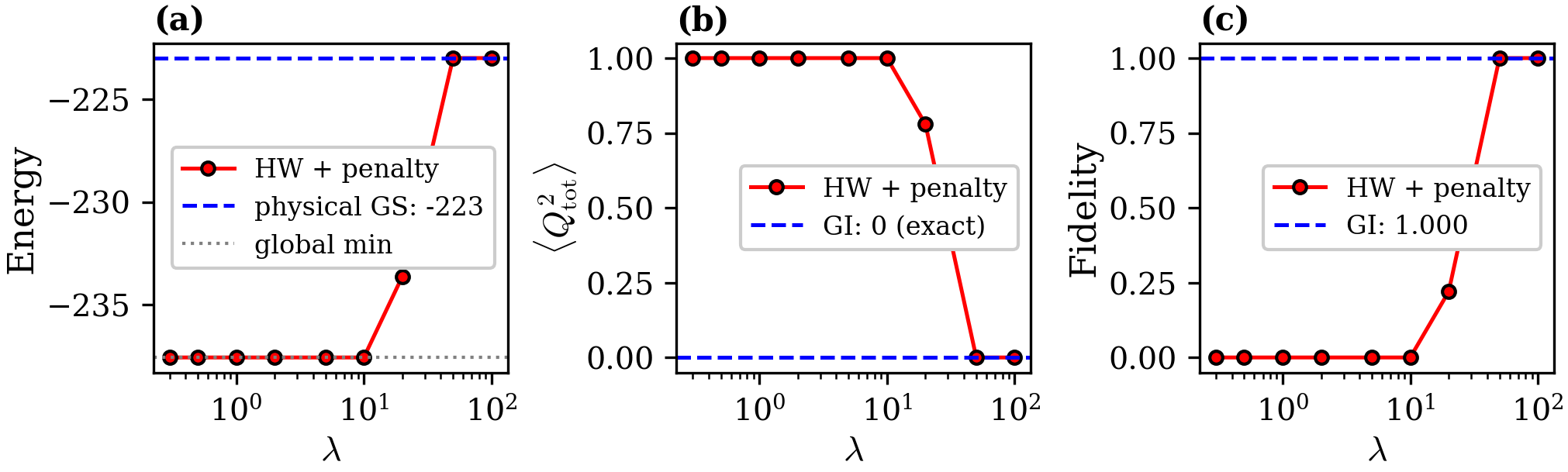}
\caption{Penalty-term sweep for the hardware-efficient ansatz at $K=-14$,
$N=2$, $L=2$. (a)~Recovered energy versus $\lambda$: it can reach the physical
ground state only for $\lambda\gtrsim16.5$, the charge-sector gap of
Eq.~\eqref{eq:lambda} (dash-dotted), and sits at the global minimum below that.
(b)~Charge-sector leakage $\expval{\Qtot^2}$ versus $\lambda$, falling from
unity to zero across the same crossover, against the identically zero GI
reference. (c)~Fidelity with the physical ground state versus $\lambda$,
rising from zero to unity at the crossover, against the GI reference.}
\label{fig:penalty}
\end{figure*}

At $K=-14$ the gap in Eq.~\eqref{eq:lambda} is $16.500$ energy units. The
behavior across the crossover is shown in Fig.~\ref{fig:penalty}. For
$\lambda\le10$ the optimizer pays $\lambda\expval{\Qtot^2}=\lambda$ but gains
$16.500$ by remaining nonphysical, so the leakage persists, as
Fig.~\ref{fig:penalty}(a) shows through the recovered energy sitting at the
global minimum. The transition occurs between $\lambda=20$ and $\lambda=50$. At
$\lambda=20$ the leakage has only partially resolved, with
$\expval{\Qtot^2}=0.78$ and a fidelity of $0.22$; by $\lambda=50$ the
charge-sector leakage of Fig.~\ref{fig:penalty}(b) has fallen to
$\expval{\Qtot^2}=0.00$ and the fidelity of Fig.~\ref{fig:penalty}(c) has
risen to $1.00$, both consistent with Eq.~\eqref{eq:lambda}.

Three limitations follow: (i)~the gap
$\lvert E_{\Qtot=0}-E_{\mathrm{global}}\rvert$ is unknown a priori, since
computing it requires solving the problem; (ii)~a large $\lambda$ distorts the
cost landscape; and (iii)~the optimizer still searches $2^{\nq}$ dimensions
rather than $\binom{2N}{N}$. The gauge-invariant construction used in this work
avoids all three limitations at a cost of a factor of two in two-qubit gates,
and we find that cost to be the cheaper of the two options at every $N$ tested,
with the advantage widening as $N$ grows. At $N=3$ the GI ansatz reaches
$0.01\%$ error with $\expval{\Qtot^2}=0.00$ and fidelity $1.00$, while the HW
ansatz gives $18.41\%$ error with $\expval{\Qtot^2}=0.63$ and a non-integer
flavor-$0$ number $\expval{N_0}=1.49$. We note that adequate restarts
($\ge\!10$) resolve a marginal-overcompleteness fidelity dip otherwise seen at
$N=3$, $L=2$, which confirms it as optimizer under-convergence rather than
near-degeneracy. Charge conservation also changes how reliably the optimizer
converges at all, which Section~\ref{sec:stability} quantifies.

\subsection{Optimization stability and convergence}\label{sec:stability}
\begin{figure*}[t]
\centering
\includegraphics[width=\textwidth]{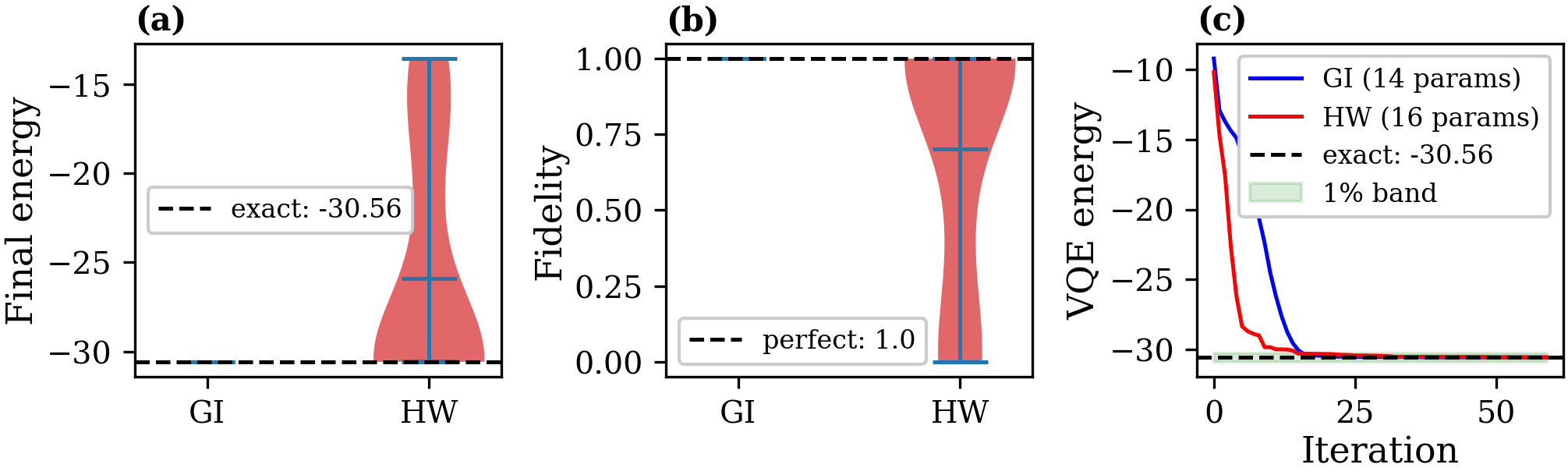}
\caption{Optimization stability and convergence ($K=0$, $N=2$, $L=2$). Across
$20$ random restarts the gauge-invariant ansatz is deterministic
($\sigma_E=1.0\times10^{-4}$, mean fidelity $0.99$) while the
hardware-efficient ansatz scatters ($\sigma_E=5.38$, mean fidelity $0.84$): the
(a)~final-energy spread and (b)~fidelity spread collapse to a line at the exact
value for the gauge-invariant ansatz but spread widely for the
hardware-efficient one. (c)~Representative convergence traces: the
gauge-invariant optimizer reaches the ground state in $\sim\!10$ iterations,
whereas the hardware-efficient optimizer plateaus in nonphysical regions before
descending to the physical ground state.}
\label{fig:stab}
\end{figure*}

Symmetry protection also stabilizes the classical optimization. Comparing the
two ansatz families at fixed $N=2$ and $L=2$ over $20$ random restarts at
$K=0$, the gauge-invariant ansatz returns an identical result every time, with
a spread of $\sigma_E=1.0\times10^{-4}$ across restarts and mean fidelity
$0.99$. The six-dimensional physical sector presents a single dominant basin
that COBYLA locates deterministically. The unconstrained ansatz, searching $16$
dimensions of which only $37.50\%$ are physical, scatters across restarts
($\sigma_E=5.38$, mean fidelity $0.84$). Roughly one restart in five is trapped
in a nonphysical minimum even at $K=0$, where no charge-sector leakage is
energetically possible; the final-energy spread of Fig.~\ref{fig:stab}(a) and
the fidelity spread of Fig.~\ref{fig:stab}(b) both collapse to a line for the
gauge-invariant ansatz and spread widely for the hardware-efficient one. The
convergence traces in Fig.~\ref{fig:stab}(c) show the mechanism: the
gauge-invariant optimizer reaches the ground state in about ten iterations,
whereas the hardware-efficient optimizer spends tens of iterations exploring
nonphysical regions before entering $\Hphys$.

Sections~\ref{sec:express} and~\ref{sec:symmetry} have treated expressibility
and trainability as separate obstacles. They are in fact controlled by the same
property of the ground state, which Section~\ref{sec:entangle} identifies.

\section{Entanglement as the Unifying Diagnostic}\label{sec:entangle}
The half-system von~Neumann entropy $S_{A|B}=-\mathrm{Tr}(\rho_A\ln\rho_A)$,
where $\rho_A$ is the reduced density matrix of one half of the lattice,
connects the three obstacles. It is computed here by Schmidt decomposition of
the exact ground state. The point is not that we compute an entanglement
entropy, which is routine, but that the single quantity $S_{A|B}$ accounts for
all three scaling obstacles at once, as the rest of this section develops. The
entropy is a sharper diagnostic than the quantum mutual information used in
tomographic hardware studies~\cite{melzer2025}, since $S(\rho_A)=0$ if and only
if the bipartition is unentangled, and it is obtained noise-free from the exact
state.

\begin{figure*}[t]
\centering
\includegraphics[width=\textwidth]{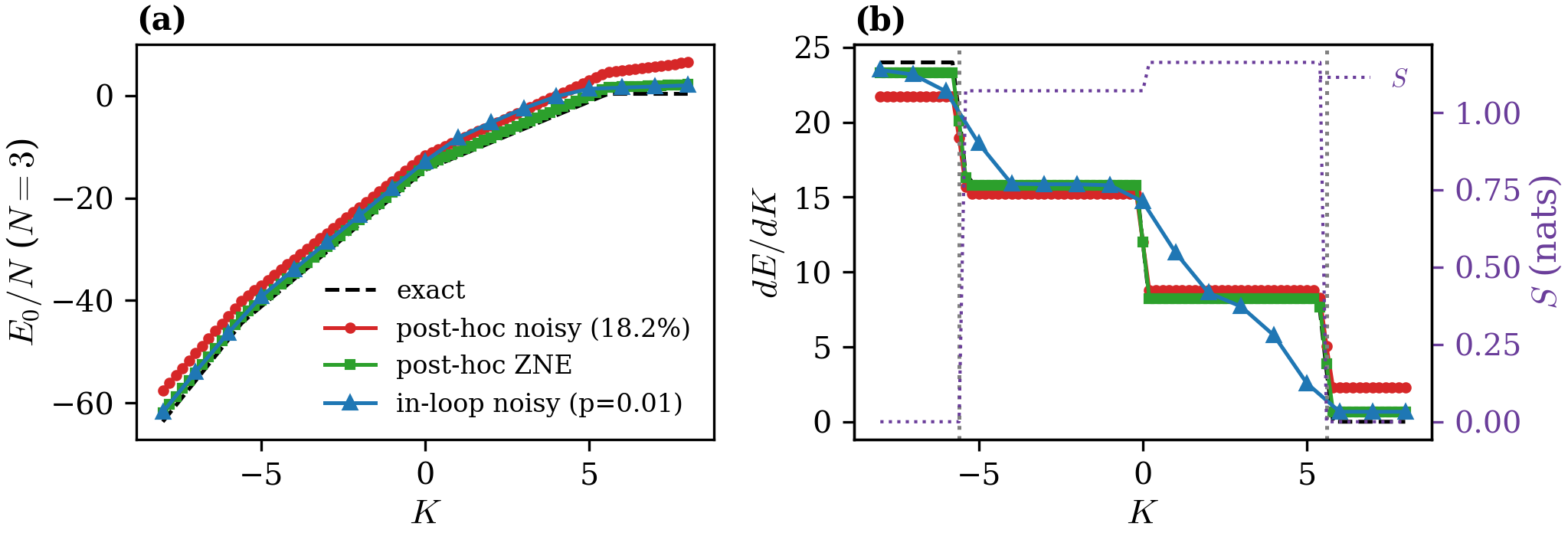}
\caption{Hardware noise on the $N=3$ model ($p=0.01$, SPAM $\varepsilon=0.005$).
(a)~Ground-state energy $E/N$ versus $K$ under the post-hoc protocol, the
in-loop protocol, and post-hoc ZNE, against the exact baseline. (b)~Slope
$dE/dK$ showing the phase-boundary kink at $\lvert K\rvert\approx5.6$,
suppressed post hoc by $19.00\%$ at $L=2$, as shown, and by $26.77\%$ at
$L=3$, against $16.16\%$ in loop at $L=3$; the half-system
entropy $S$ (dotted, right axis) forms a plateau across the central phase and
collapses to zero at the boundary. Curve coding in panel (b) is shared with
panel (a).}
\label{fig:noise}
\end{figure*}

The structure of $S(K)$ is a plateau followed by a collapse (see
Fig.~\ref{fig:noise}(b), right axis). Across the central phase the ground state
is strongly entangled, with $S$ exceeding $\ln 2$ for every interior $K$ and
reaching a maximum $S_{\max}\approx1.160$~nats near $K=0$. It falls to zero at
the phase boundary, where the state polarizes into a near-product configuration
of definite flavor number. This central-phase entanglement ties together the
first two obstacles. A ground state of half-system Schmidt rank $r$ requires an
ansatz able to generate rank at least $r$, which is why the interior demands
$L=2$ (Section~\ref{sec:express}). The same high rank produces the rugged
landscape that slows the classical optimization (Section~\ref{sec:symmetry}).
The phase boundary plays a complementary role. There the entropy collapses, but
the energy develops the slope discontinuity $dE/dK$ that marks the first-order
transition, and Section~\ref{sec:noise} shows that hardware noise degrades
precisely that discontinuity.

The region where the ansatz is under the most strain is not the boundary
itself, and the reason is not entanglement. Across $K\in[-16,16]$ at $N=3$ the
$L=2$ ansatz tracks the exact energy with a mean absolute error of $0.924$
energy units, but the deviation is confined to $K\in[1.0,5.0]$, rising to
$7.050$ energy units at $K=3.0$ and falling back to the $10^{-8}$ level by
$K=6.0$. It is negligible at every $K<1.0$, so the failure is one-sided, as
the chemical-potential convention of Section~\ref{sec:qubitham} requires.
Section~\ref{sec:boundary_express} showed that the deviation is a
representation failure that a third layer removes, and $L=3$ holds the same
$0.006$-energy-unit optimizer floor at every $K$ in that window, not only at
the single point tested there. The entropy is at its minimum at the boundary,
so a near-product state is not by itself easy to prepare; what matters is the
flavor-sector structure of Section~\ref{sec:sector}, which leaves the $L=2$
manifold unable to contain the ground state of one sector. Entanglement
therefore controls expressibility and trainability throughout the central
phase, the sector structure controls them where the flavor occupation changes,
and hardware noise degrades detection of the transition. The entropy locates
the first of these effects and the sector structure of
Section~\ref{sec:sector} locates the second.

\emph{Finite-size scaling.} Taking the leading correction to be linear in
$1/N$, so that $E/N=E_\infty+c_1/N+O(N^{-2})$ with $E_\infty$ the
thermodynamic-limit energy per site, a fit over $N=2$--$6$ gives an indicative
estimate $E/N\approx-18.650$ at $K=0$. A quadratic fit extrapolates to
$-23.810$, but with five non-monotone data points, whose non-monotonicity is
the even-odd staggering of Fig.~\ref{fig:foundation}(c), the quadratic result
is an overfitting artifact: the quadratic curve already diverges below $N=3$.
The linear estimate is preferred as the lowest-order physically motivated
correction. VQE per-site energies coincide with the exact values at $N=2$ and
$3$ to better than $0.05\%$, but the $8.29\%$ error at $N=4$, $L=2$ would shift
the estimate by several units, so reliable extrapolation at $N=4$ requires
$L\ge5$. The entropy identifies where the physics is hardest to represent. What
remains is to ask whether the hardware can execute the circuit that the
representation requires.

\section{Obstacle III: Hardware Noise and Mitigation}\label{sec:noise}
The third of the three obstacles to scaling VQE, listed in
Section~\ref{sec:model}, is hardware noise, and it is the one that binds at
$N\ge4$. It is also the obstacle whose measured size depends most strongly on
how the measurement is set up, so we treat the protocol itself as part of the
result. Section~\ref{sec:protocol} defines the two protocols we compare.
Section~\ref{sec:shots} isolates the irreducible cost of finite measurement.
Section~\ref{sec:depol} adds gate errors and measures their effect on the phase
boundary under both protocols. Section~\ref{sec:zne} tests error mitigation.
Section~\ref{sec:proj} measures the budget at $N=4$.

\subsection{Two noise protocols}\label{sec:protocol}
Studies of variational algorithms under noise adopt one of two protocols, and
they are not equivalent. We define both here and use both throughout. In the
\emph{post-hoc} protocol, the ansatz is optimized in the absence of noise. The
converged state is then embedded in the full Hilbert space, a cumulative
two-qubit depolarizing channel with surviving fraction
$\wpass=(1-p)^{n_{\mathrm{CX}}}$ is applied, where $p$ is the per-CNOT
depolarizing probability and $n_{\mathrm{CX}}$ the CNOT count, an independent
per-qubit bit-flip channel with state-preparation-and-measurement (SPAM) error
$\varepsilon=0.005$ follows, and $\expval{W}=\mathrm{Tr}(W\rho)$ is read out.
The optimizer never sees the noise. In the \emph{in-loop} protocol, a two-qubit depolarizing channel of probability $p$ is inserted
after every CNOT, with the same SPAM channel before readout, and the cost function evaluated at every
optimizer step is $\expval{W}=\mathrm{Tr}(W\rho)$ on the noisy density matrix.
The optimizer is free to move the parameters in response to the channel.

The distinction matters because the two protocols make different predictions
for the same observable. Under the post-hoc protocol the state is a fixed
convex mixture of the noiseless state and the maximally mixed state of the
physical sector, so
every derivative of the energy with respect to an external parameter is
contracted by the passive fraction $\wpass$. Under the in-loop protocol no such statement holds. The two protocols
therefore differ in two respects at once, in where the noise acts, as a global
contraction or as per-gate local channels, and in whether the optimizer sees
it. Section~\ref{sec:depol} separates the two with a fixed-parameter control
in the local channel.

All the in-loop results reported in Sections~\ref{sec:depol}
to~\ref{sec:proj} use the derivative-free COBYLA optimizer~\cite{powell1994} on
a density-matrix simulation, validated against the noiseless energies of
Table~\ref{tab:vqe} at $p=0$ and $\varepsilon=0$ to within $3\times10^{-4}$
energy units. Chemical-potential scans use warm-start continuation, in which
the optimization at each $K$ is seeded from the converged parameters at the
preceding $K$; the sweep is performed in both directions and the lower energy
is retained at each point. Without warm-start continuation, the optimizations
at neighboring $K$ land in different local optima and the resulting curve is
not differentiable.

\subsection{Finite-shot measurement budget}\label{sec:shots}
Before modeling gate errors we isolate the statistical cost of finite
measurement, which is present even on a perfect device. Expanding
$W=\sum_i c_iP_i$ into Pauli strings $P_i$ with coefficients $c_i$, and
sampling each string with $n_{\mathrm{shots}}=100$ from the exact ground state,
gives a shot-noise energy with standard deviation
$\sigma_E\le\sum_i\lvert c_i\rvert/\sqrt{n_{\mathrm{shots}}}$.
Table~\ref{tab:shot} reports the resulting estimates at $N=2$. The additional
relative error is $0.29\%$ at $K=-14$ and $0.15\%$ at $K=0$, consistent with
the bound. The $304.00\%$ figure at $K=+10$ is the denominator artifact
described in Section~\ref{sec:leakage}: the exact energy $+1.000$ lies near
zero on the scale of $W$, whose range is $-223.000$ to $+1.000$, so the
absolute error of $3.040$ units is in fact small.

Measurement noise alone is therefore sub-percent at $100$ shots, and the
dominant hardware error is gate infidelity, which is modeled in
Section~\ref{sec:depol}. We note that $W$ contains $8$ nontrivial Pauli terms
at $N=2$ and $O(N^2)$ in general, dominated by the long-range electric $ZZ$
couplings, and that the bound above measures each term independently.
Commuting-set grouping, in which the $Z$-diagonal terms share a single
measurement basis, would lower the budget further, so the reported figure is a
conservative upper bound.

\begin{table}[b]
\caption{Finite-shot energy estimates at $100$ shots per Pauli string ($N=2$,
$x=16$). Energies are in the dimensionless units of Eq.~\eqref{eq:W}.}
\label{tab:shot}
\centering
\begin{tabular}{crrrr}
\toprule
$K$ & exact & shot mean & $\sigma_E$ & rel.\ err.\ (\%)\\
\midrule
$-14$ & $-223.000$ & $-223.640$ & $1.600$ & $0.29$\\
$0$   & $-30.564$  & $-30.520$  & $0.180$ & $0.15$\\
$+10$ & $+1.000$   & $-2.040$   & $1.600$ & $304.00^\ast$\\
\bottomrule
\end{tabular}
\\[2pt]
{\footnotesize $^\ast$Denominator artifact; the absolute error of $3.040$ units
is small. See Section~\ref{sec:leakage}.}
\end{table}

\subsection{Depolarizing noise and the phase boundary}\label{sec:depol}
We now compare the two protocols defined in Section~\ref{sec:protocol} on the
observable that matters physically, which is the slope discontinuity $dE/dK$ at
the first-order transition. The comparison is made at $N=3$ with $L=3$, the
layer count Section~\ref{sec:boundary_express} showed to be necessary in this
region, on a grid of $13$ chemical potentials spanning $K\in[4.0,7.0]$. At
$L=3$ the circuit carries $n_{\mathrm{CX}}=30$ CNOTs, giving a cumulative
infidelity $1-(0.99)^{30}=26.00\%$ at $p=0.01$ and a passive fraction
$\wpass=(1-p)^{30}(1-2\varepsilon)=0.732$.

The in-loop energies are not a contracted copy of the noiseless ones. They are
displaced by a nearly constant offset with the slope left largely intact.
Table~\ref{tab:protocol} collects the branch fits. On the ordered branch the
in-loop slope is $7.672$ against the noiseless $8.000$, a ratio of $0.959$, and
the offset between the two curves is $7.093$ with a standard deviation of
$0.124$ across the branch. The passive protocol predicts a slope ratio of
$\wpass=0.732$ instead. The residual slope of $0.965$ on the saturated branch,
where the exact energy is flat, is not a failure of the ansatz. The maximally
mixed reference energy grows with $K$ at a rate
$dE_{\mathrm{mix}}/dK=12.000$ at $N=3$, since each of the three flavor-$0$
sites contributes $\nu_0/2$ with $\nu_0=2\sqrt{x}K$, so any admixture of that
reference produces an upward slope where the exact curve has none.

\begin{table}[b]
\caption{Branch fits of $E(K)$ at $N=3$, $L=3$, $x=16$, on the $13$-point grid
$K\in[4.0,7.0]$. The ordered branch has flavor-$0$ occupation $N_0=1$ and the
saturated branch $N_0=0$. The in-loop protocol largely preserves the slope; the
passive protocol predicts contraction by $\wpass=0.732$.}
\label{tab:protocol}
\centering
\begin{tabular}{lrrr}
\toprule
 & noiseless & in-loop, $p=0.01$ & ratio\\
\midrule
ordered branch, $dE/dK$   & $8.000$ & $7.672$ & $0.959$\\
saturated branch, $dE/dK$ & $0.000$ & $0.965$ & ---\\
offset, ordered branch    & ---     & $7.093(124)$ & ---\\
\bottomrule
\end{tabular}
\end{table}

That observation gives a protocol-independent way to read the data. Writing the
in-loop slope on each branch as
$\weff\,(dE_{\mathrm{exact}}/dK)+(1-\weff)(dE_{\mathrm{mix}}/dK)$ and solving
for the retained fraction $\weff$ gives $\weff=1.082$ from the ordered branch
and $\weff=0.920$ from the saturated branch. Both lie far above the passive
value $\wpass=0.732$, and neither requires a fitted parameter. In-loop
optimization therefore retains between $91.96$ and $108.21\%$ of the
energy-derivative signal at a cumulative infidelity of $26.00\%$, where passive
contraction would retain $73.20\%$.

The ordered-branch value exceeds unity, and we report it as solved rather than
clipping it. The mixing model writes the in-loop slope as a convex combination
of $dE_{\mathrm{exact}}/dK=8.000$ and $dE_{\mathrm{mix}}/dK=12.000$, both of
which the measured in-loop slope of $7.672$ falls below, so no $\weff$ in
$[0,1]$ reproduces it. The excess is not fit error: the in-loop slope sits
$0.328$ energy units under the noiseless one, four times the largest
branch-fit residual of $0.082$. What the two-reference model cannot capture is
that the in-loop optimizer does not interpolate between the noiseless state and
the maximally mixed one; it finds a different variational minimum at each $K$,
and on this branch that minimum tracks the exact curve with a slope slightly
shallower than either reference. The saturated branch, where the measured slope
does lie between the two references, returns a value inside the physical range.
Both branches place the retained signal far above the passive prediction.

A control separates the two differences between the protocols. We optimize
noiselessly at each $K$, keep every restart that reaches the exact energy to
within $0.05$ units, and evaluate those fixed parameters in the same local
channel at $p=0.01$ without re-optimization. Equally good noiseless optima give
noisy energies that differ by $0.5$ to $3.3$ units at the same $K$, so we report
the distribution over $20000$ random selections of one such optimum per $K$.
The kink suppression has median $9.79\%$ with a $5$ to $95\%$ range of
$-12.00$ to $36.85\%$, and the retained fractions have medians $1.021$ on the
ordered branch and $0.949$ on the saturated branch. The in-loop values of
$16.16\%$, $1.082$ and $0.920$ lie inside these distributions. Optimizing
inside the channel therefore gives no measurable advantage over evaluating a
noiselessly optimized state in the same channel. What separates both from the
passive estimate is the noise model: per-gate local channels degrade the
transition signature far less than the global contraction $\wpass$ that the
post-hoc formula assumes.

The consequence for the transition is direct. Taking the kink magnitude as the
difference between the ordered- and saturated-branch slopes, the noiseless
value is $8.000$ and the in-loop value is $6.707$, a suppression of $16.16\%$.
The global contraction applied to the same $L=3$ circuit gives $26.77\%$, the
passive-contraction fraction $1-\wpass$, as it must. The location of the
transition is preserved under both protocols; only its magnitude is reduced, and under local noise it is reduced by less
than the global contraction predicts, with or without re-optimization.

We note one methodological point that the reader will need if reproducing the
calculation. A one-directional continuation sweep cannot cross a first-order
boundary, because the optimal parameters change discontinuously there. Sweeping
in both directions produces two branches that disagree only in the window
$K\in[4.00,5.50]$, which spans the approach to the transition at
$\lvert K\rvert\approx5.6$. The disagreement is the signal, not an artifact. Error mitigation is treated in
Section~\ref{sec:zne}.

\subsection{Zero-noise extrapolation}\label{sec:zne}
\begin{figure}[t]
\centering
\includegraphics[width=\columnwidth]{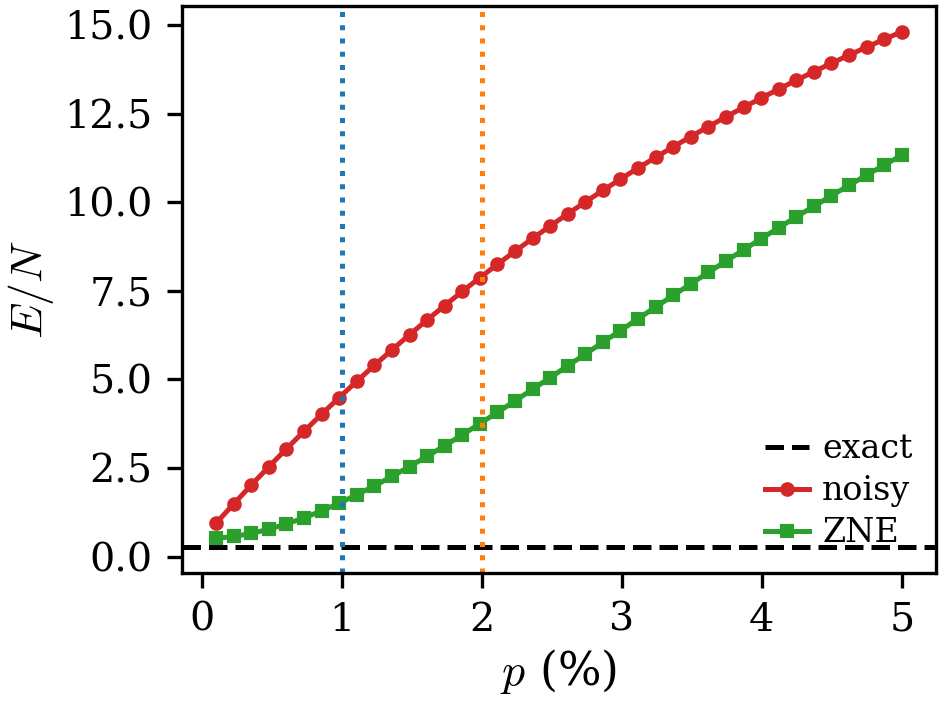}
\caption{Ground-state energy $E/N$ at the $N=3$ phase boundary versus per-CNOT
depolarizing probability $p$, unmitigated and after first-order zero-noise
extrapolation under the post-hoc protocol. The residual is the gap between the
mitigated curve and the exact baseline (dashed). It grows smoothly and
monotonically, because linear Richardson extrapolation cannot match the
exponential depolarizing decay. The in-loop residuals of Table~\ref{tab:zne}, computed with per-gate local
channels, are between four and nine times smaller at the same
values of $p$. The exact baseline lies at $E/N=0.28$, and the mitigated curve
uses a noise-scaling factor $\lambda=3$.}
\label{fig:robust}
\end{figure}

Error mitigation is the standard response to the bias measured in
Section~\ref{sec:depol}. We fold the circuit to a noise-scaling factor
$\lambda=3$ and apply first-order Richardson zero-noise
extrapolation~\cite{temme2017,li2017,giurgica2020},
$E_{\mathrm{mit}}=\tfrac{3}{2}E(1)-\tfrac{1}{2}E(3)$. Here $\lambda$ is the
conventional ZNE noise-amplification factor and is unrelated to the penalty
strength of Section~\ref{sec:penalty}. Under the post-hoc protocol, ZNE leaves
a residual at the boundary that grows smoothly and monotonically with $p$:
$3.80$ energy units at $p=1.00\%$, $10.50$ at $p=2.00\%$, and $33.00$ at
$p=5.00\%$, as shown in Fig.~\ref{fig:robust}. The cause is a model mismatch.
The depolarizing energy is exponential in $\lambda$, as
$E(\lambda)\propto \wpass^{\lambda n_{\mathrm{CX}}}$, whereas Richardson
extrapolation assumes linearity, so the extrapolation systematically
overshoots.

Under the in-loop protocol, in which each folded circuit is optimized
separately in the local channel, the residuals are $0.84$, $1.20$ and $4.02$
energy units at the same three noise levels. Table~\ref{tab:zne} collects both
columns. They differ in the noise model as well as the protocol, for the reason
set out in Section~\ref{sec:depol}, so the factor of four to eight between them
is not an effect of re-optimization alone, and the fixed-parameter control of
that section indicates that it is mostly the model. In-loop extrapolation is
also not the same estimator as the post-hoc one, since the optimizer produces a
different variational minimum at each $\lambda$.

\begin{table}[t]
\caption{Zero-noise extrapolation residual at the $N=3$ phase boundary, in the
dimensionless energy units of Eq.~\eqref{eq:W}, under the two protocols of Section~\ref{sec:protocol}. The post-hoc column
uses the global contraction and the in-loop column per-gate local channels.}
\label{tab:zne}
\centering
\begin{tabular}{crr}
\toprule
$p$ (\%) & post-hoc & in-loop\\
\midrule
$1.00$ & $3.80$  & $0.84$\\
$2.00$ & $10.50$ & $1.20$\\
$5.00$ & $33.00$ & $4.02$\\
\bottomrule
\end{tabular}
\end{table}

The practical conclusion is that the commonly quoted limit of $p\approx2.00\%$
for first-order ZNE reflects the global contraction model. Under per-gate local
noise at $N=3$ the residual is $0.84$ units at $p=1.00\%$, comparable to the
magnitude of the exact boundary energy, and $4.02$ units at $p=5.00\%$. Whether the
advantage survives to larger systems is the subject of
Section~\ref{sec:proj}.

\subsection{Simulated budget at $N=4$}\label{sec:proj}
The $N=4$ circuit requires $L_{\min}=5$ and $2L_{\min}(2N-1)=70$ CNOTs, giving
a cumulative infidelity $1-(0.99)^{70}=50.50\%$ at $p=1.00\%$. We measure the
consequence directly rather than projecting it, by running in-loop VQE at three
chemical potentials and comparing against a noiseless control at the same three
points, with an identical restart budget of four restarts and a maximum of
$2000$ iterations. The three points, $K=0.00$, $1.25$ and $2.50$, span the
low-$K$ region well inside the central phase; they are not boundary points, the
terminal $N=4$ crossing lying at $K\approx6.36$ (Section~\ref{sec:sector}). The
control is necessary because COBYLA reports a failure to converge on
essentially every restart at $N=4$, including in a noiseless run that
reproduces the reference energy to $3\times10^{-4}$. Its convergence flag
therefore carries no information at this system size, and the error against the
exact ground state is the only meaningful signal.

Table~\ref{tab:n4} gives the comparison. The noiseless control reaches a mean
error of $0.12\%$, so the restart budget is adequate. At $p=1.00\%$ the same
circuit reaches $25.69$ to $52.81\%$, and at $p=2.00\%$ it reaches $43.23$ to
$88.69\%$. The degradation is therefore attributable to noise and not to the
optimization budget. Mitigation does not rescue the result. Folding to
$\lambda=3$ would require $210$ CNOTs and a cumulative infidelity
$1-(0.99)^{210}\approx88.00\%$, which renders the folded circuit
informationally indistinguishable from the maximally mixed state. The favorable
in-loop mitigation of Section~\ref{sec:zne} therefore does not extend to $N=4$
at present gate fidelities, and hardware noise is the binding constraint at
that size.

The circuit depths and gate counts quoted above assume that the $(\nq-1)$
exchange gates of a layer act on neighbouring qubits, which is the connectivity
of the trapped-ion register of Ref.~\cite{melzer2025}. We checked whether that
assumption survives on a fixed-coupling superconducting architecture by
transpiling the ansatz onto a $127$-qubit heavy-hex map at the highest
optimization level, over $20$ random routing seeds. The two-qubit gate count is
unchanged in every case: $20$, $30$ and $70$ gates at $(N,L)=(3,2)$, $(3,3)$
and $(4,5)$ respectively, with no SWAP inserted at any seed. The reason is that
a heavy-hex lattice contains long chains of degree-two qubits between its
branch points, onto which a strictly one-dimensional circuit of six or eight
qubits embeds without routing. Only the decomposed depth rises, by roughly a
factor of two, through the translation of each exchange gate into the native
two-qubit basis. The gate-count results of Section~\ref{sec:condition} are
therefore not specific to trapped-ion connectivity at the sizes studied here.

The per-gate error rate, however, is specific to the platform, and the noise
budget of this section is not. At a two-qubit error of $1.17\times10^{-3}$, an
order of magnitude below the $p=1.00\%$ used throughout
Sections~\ref{sec:depol} to~\ref{sec:proj}, the cumulative infidelity at
$N=3$, $L=3$ falls from $26.00\%$ to $3.45\%$, and at $N=4$, $L=5$ from
$50.50\%$ to $7.87\%$.
% Hardware: job daoeuotr85ps73ff6a70 (ibm_marrakesh). Repeats: daofnv78gn2s739nqjlg,
% daofnvopqrnc7399u9ig, daonh9lr85ps73ffhqjg. Estimator-primitive runs excluded
% (server dropped two Y-Z-Y terms; see results/ibm_N3_L2_K0_gap_localize.json).
A reported per-gate error does not, however, fix the effective error of a
complete circuit. We evaluated the $N=3$, $L=2$ circuit at $K=0$ on a
$156$-qubit superconducting processor, with $20$ native two-qubit gates and
$8192$ shots per measurement basis. The measured energy lies $15.70\%$ above
the exact value, and $11.20\%$ above it after correcting the counts for the
calibrated readout error of the six qubits used. Within the noise model of
Section~\ref{sec:depol} these correspond to effective two-qubit errors of
$0.76\%$ and $0.57\%$, two to three times the reported median of $0.29\%$.
Three repeat evaluations on the same qubits over the following ten hours gave
$17.35$ to $18.15\%$, so the first run is, if anything, the favorable one. At
the readout-corrected rate the $N=4$, $L=5$ circuit accumulates a cumulative
infidelity of $32.84\%$, so the identification of $N=4$ as noise-limited holds
at the error rate this device delivers, and the location of the frontier moves
with the device.
What the expressibility condition fixes is the gate count a given lattice size
demands; what the device fixes is whether that count is affordable.

\begin{table}[b]
\caption{In-loop VQE at $N=4$, $L=5$, $x=16$, against a noiseless control at
matched chemical potentials with an identical four-restart budget. The three
$K$ values lie inside the central phase, not at the $N=4$ boundary. Relative
error against the exact charge-neutral ground state.}
\label{tab:n4}
\centering
\begin{tabular}{crrr}
\toprule
$K$ & noiseless (\%) & $p=1.00\%$ & $p=2.00\%$\\
\midrule
$0.00$ & $0.08$ & $25.69$ & $43.23$\\
$1.25$ & $0.05$ & $34.02$ & $57.11$\\
$2.50$ & $0.22$ & $52.81$ & $88.69$\\
\bottomrule
\end{tabular}
\end{table}

\section{Discussion}\label{sec:discussion}
For the charge-neutral two-flavor Schwinger model studied here, VQE accuracy is
controlled by a single dimensionless ratio, $p/d$, of the variational parameter
count to the physical-sector dimension. When $p/d<1$ the ansatz manifold cannot
contain the ground state and no number of restarts helps. When $p/d\ge1$ the
manifold is in principle complete, and the remaining difficulties are landscape
ruggedness, which is manageable with a handful of restarts, and circuit
ordering and connectivity, which must be checked per system size. The ratio
$p/d$ is necessary but not sufficient, and it must be evaluated where the state
is actually being prepared: at $N=3$ the layer count that suffices at $K=0$
leaves a $74.38\%$ error near the first-order boundary, and one additional
layer removes that error at a cost of $7.80$ percentage points of cumulative
infidelity.

Charge conservation protects against two otherwise serious obstacles at once.
It slows the decay of the gradient variance by nearly a factor of four
relative to an unconstrained circuit, and it forbids
the charge-sector leakage that makes an unconstrained ansatz report converged
but unphysical energies; measuring $\expval{\Qtot^2}$ is the cheap diagnostic
that detects the leakage. The binding constraint at $N\ge4$ is therefore
neither trainability nor expressibility in principle, but hardware noise: the
depth-$30$, $70$-CNOT circuit needed for $N=4$ incurs a $50.50\%$ cumulative
infidelity, and a noiseless control at matched chemical potentials confirms
that the resulting $25.69$ to $52.81\%$ errors are caused by the noise and not
by the optimization budget.

How that noise is modeled changes the answer by a large factor, which is the
methodological result of this work. Per-gate local noise retains between
$91.96$ and $108.21\%$ of the energy-derivative signal at $26.00\%$ cumulative
infidelity, where the global contraction commonly used for post-hoc estimates
retains $73.20\%$, and a fixed-parameter control shows that re-optimizing
inside the channel adds nothing measurable on top of that. The noise model is
therefore a necessary part of any reported noise budget for a variational
algorithm, and the global contraction is a conservative bound rather than an
estimate. Single-realization comparisons are also unreliable at this level:
equally good noiseless optima differ by up to $3.3$ energy units once noise is
applied.

\emph{Limitations.} Five limitations should be noted. (i)~The noise model is
incoherent, comprising depolarizing and SPAM channels only, so coherent gate
errors, crosstalk and drift are not captured; the budgets are quoted at
$p=1.00\%$ and scale with the two-qubit error rate of the device, as
Section~\ref{sec:proj} shows.
(ii)~The optimizer is the derivative-free COBYLA, which converges faster than
the SPSA used on hardware~\cite{spall1992}. (iii)~The hardware-efficient
comparison uses a fixed nearest-neighbor ladder, so the charge-sector-leakage
magnitudes are representative rather than worst-case. (iv)~The $p/d$ collapse
places points on both sides of unity at $N=2$, $3$ and $4$; at $N=5$ and $6$
the crossing lies at $L=14$ and $L=41$, beyond the depths we ran, so those two
sizes constrain the undercomplete side only. (v)~The retained fraction $\weff$ is solved from a two-reference mixing model
that assumes a global contraction. Under per-gate local noise the ordered-branch
slope falls outside it for in-loop and fixed-parameter states alike, so values
above unity signal that mismatch rather than a physical fraction.

Establishing the generality of the $p/d$ collapse beyond the setting studied
here is left to future work. The open cases are other values of the coupling
$x$, open versus periodic boundary conditions, alternative intra-layer gate
orderings and connectivities, and other fermion discretizations. The ordering
dependence at $L=1$ documented in Section~\ref{sec:collapse} makes a controlled
brick-wall against sequential comparison at fixed $p/d$ the most immediate of
these, and it is also what would settle the reachable-set explanation offered
in that section. We do not expect any of the open cases to alter the
expressibility condition or the symmetry-protection mechanism.

\emph{Future directions.} Crossing $p/d=1$ at $N\ge4$ with lower depth
motivates ADAPT-VQE~\cite{grimsley2019}, in which the circuit is grown greedily
from a gate pool, and symmetry-compressed ans\"atze~\cite{gard2020}. Mitigation
beyond first-order ZNE, whether higher-order extrapolation, learning-based
extrapolation, or probabilistic error cancellation, is required for
quantitative results at $N=4$. Extending the chemical-potential analysis to
$K$-dependent penalty schedules, and extending the model to real-time dynamics
and the $\theta$-term~\cite{funcke2020}, are natural next steps.
Finite-temperature spectral functions and energy-energy correlators have
already been computed for the single-flavor model via real-time
simulation~\cite{barata2025}, which illustrates the physical targets that
motivate scaling VQE to larger $N$. The threshold at which quantum hardware
outperforms the classical tensor network benchmarks now available for this
model~\cite{schwaegerl2025} lies in the confinement and string-breaking
regimes, which are accessible only beyond NISQ-era coherence budgets.

\section{Conclusion}
We have presented a unified VQE scaling study of the two-flavor Schwinger model
from $N=2$ to $6$ staggered lattice sites, disentangling expressibility,
symmetry protection, and hardware noise within one convention and one
reproducible codebase. The expressibility condition
$L(4N-1)\gtrsim\binom{2N}{N}$ predicts when a multi-layer charge-conserving
ansatz can reach the ground state, and the energy error collapses onto a single
function of $p/d$. The condition is evaluated at a point in the phase diagram
rather than globally: at $N=3$ the layer count sufficient at $K=0$ fails
across the flavor sector that ends at the first-order boundary, where a third
layer is required.

Charge conservation provides a double protection against gradient decay and
charge-sector leakage. The latter is quantified by the penalty strength
$\lambda\ge\lvert E_{\Qtot=0}-E_{\mathrm{global}}\rvert$, derived in
Section~\ref{sec:penalty}, and detected through $\expval{\Qtot^2}$.

The half-system entanglement entropy is the common diagnostic. It is high
across the entire central phase, which is what forces the $L=2$ expressibility
threshold and the rugged optimization landscape, and it collapses at the phase
boundary, where instead the energy slope discontinuity that marks the
first-order transition is the feature most degraded by hardware noise.

Hardware noise is the binding constraint at $N\ge4$, but its size depends on
how it is modeled. Under per-gate local noise the energy-derivative signal is
retained at $91.96$ to $108.21\%$ at $26.00\%$ cumulative infidelity, against the
$73.20\%$ of a global contraction, and the first-order slope discontinuity is
suppressed by $10$ to $16\%$ rather than $26.77\%$, whether or not the optimizer
runs inside the channel. These results identify $N=3$ as the
immediately viable extension of existing trapped-ion experiments, at $L=2$ and two-qubit depth $14$ with $22$ parameters on six qubits away from the boundary, and $L=3$
within it, given trapped-ion gate fidelities: on a present superconducting
device the same $L=2$ circuit returned energies $15.70$ to $18.15\%$ above the
exact value without error mitigation. At $N=4$ the requirement is $L=5$, and the $50.50\%$ cumulative
infidelity of that circuit is not recoverable by folded extrapolation at
present gate fidelities.

\section*{Reproducibility}
All energies, gradients, ansatz comparisons, penalty sweeps, shot-noise budgets,
and depolarizing and ZNE results were generated from a single statevector and
density-matrix codebase in one chemical-potential convention. The code, the raw
output of every sweep reported here, and the scripts that regenerate every
figure are publicly available at \url{https://github.com/0Shunya0/VQE_LGT}.
Each figure regenerates from the committed data, with the printed diagnostics
matching the values reported here. The occupation convention of
Eq.~\eqref{eq:Welec} is asserted as a unit test against the analytic $N=2$
electric term, and the full-space and sector-projected Hamiltonians are checked
to return the same expectation value on the exact ground state at several
chemical potentials, which fixes the relative sign between the two diagonal
terms. The in-loop density-matrix evaluator was cross-validated against an
independent implementation and against exact statevectors, and reproduces the
noiseless energies of Table~\ref{tab:vqe} to within $3\times10^{-4}$ energy
units at $p=0$ and $\varepsilon=0$. The exact-diagonalization benchmarks
reproduce the published values of~\cite{melzer2025} exactly at $K=-14$ and
$K=+10$, and to $0.136$ energy units at $K=0$, as tabulated in
Table~\ref{tab:verify}, and are consistent with the classical tensor network
computations of~\cite{schwaegerl2025} on the same model. The hardware
evaluations of Section~\ref{sec:proj} were run on ibm\_marrakesh as IBM Quantum
jobs daoeuotr85ps73ff6a70, daofnv78gn2s739nqjlg, daofnvopqrnc7399u9ig and
daonh9lr85ps73ffhqjg; their raw counts are included in the repository.

\section*{Acknowledgment}
The authors thank the PES University Department of Electronics and Communication
Engineering for computational support. We acknowledge the use of IBM Quantum
services for this work. The views expressed are those of the authors, and do not
reflect the official policy or position of IBM or the IBM Quantum team.

\bibliography{references}

\end{document}